\documentclass[12pt]{article}
\usepackage{epsfig}
\usepackage{graphicx}
\usepackage{a4}
\usepackage{latexsym}
\usepackage{cite}

\usepackage{color}
\usepackage{colordvi}

\usepackage{pslatex}
\usepackage[latin1]{inputenc}
\usepackage[T1]{fontenc}

\def\spinorbra#1#2{\langle#1#2\rangle}

\newcommand{\gsim}{\raisebox{-0.07cm}{$\:\:\stackrel{>}{{\scriptstyle
 \sim}}\:\: $} }
\newcommand{\lsim}{\raisebox{-0.07cm}{$\:\:\stackrel{<}{{\scriptstyle
 \sim}}\:\: $} }
\def\slash#1{\rlap{\hbox{$\mskip 1 mu /$}}#1}

\begin{document}
\setlength{\parskip}{0.2cm}
\setlength{\baselineskip}{0.55cm}

\begin{titlepage}
\noindent
DESY 08-019 \hfill {\tt arXiv:0803.0457v2}\\[1mm]
February 2008 \\
\vspace{1.8cm}
\begin{center}
\Large
{\bf Hard QCD at hadron colliders} \\
\vspace{2.2cm}
\large
S. Moch\\
\vspace{1.4cm}
\normalsize
{\it Deutsches Elektronensynchrotron DESY \\
\vspace{0.1cm}
Platanenallee 6, D--15738 Zeuthen, Germany}\\
\vspace{5.0cm}
\large
{\bf Abstract}
\vspace{-0.2cm}
\end{center}
We review the status of QCD at hadron colliders with emphasis on precision predictions 
and the latest theoretical developments for cross sections calculations to higher orders.
We include an overview of our current information on parton distributions and 
discuss various Standard Model reactions such as $W^\pm$/$Z$-boson, Higgs
boson or top quark production.
\vfill
\end{titlepage}

%
% -----------------------------------------------------------------------------
%
\section{Introduction}
\label{sec:intro}

Historically, hadron colliders have explored elementary particle physics at the energy frontier 
the motivation being the discovery of new particles through direct production.
This has been the case for the SppS at CERN leading to the discovery of the
weak vector bosons as well as for Tevatron at Fermilab currently operating at a 
center-of-mass energy ${\sqrt{S}}=1.96$~TeV with the discovery of the top quark.
Shortly the Large Hadron Collider LHC at CERN with ${\sqrt{S}}=14$~TeV will commence operation which
will realize a major leap forward in collision energy.
Being long awaited the machine will allow access to the mechanism 
of electro-weak symmetry breaking, to search for the Higgs boson and,
hopefully, it will open new avenues to test many proposed extensions of the Standard Model.
To that end, two general purpose experiments ATLAS and CMS~\cite{atlas:1999tdr1,atlas:1999tdr2,cms:2006tdr,Ball:2007zza}
as well as two specialized one, LHCb for $B$-physics~\cite{lhcb:2003tdr}
and Alice for heavy-ion physics~\cite{alice:2005tdr}, have been installed.

\begin{figure}[htb]
  \begin{center}
    \includegraphics[width=7.5cm,angle=0]{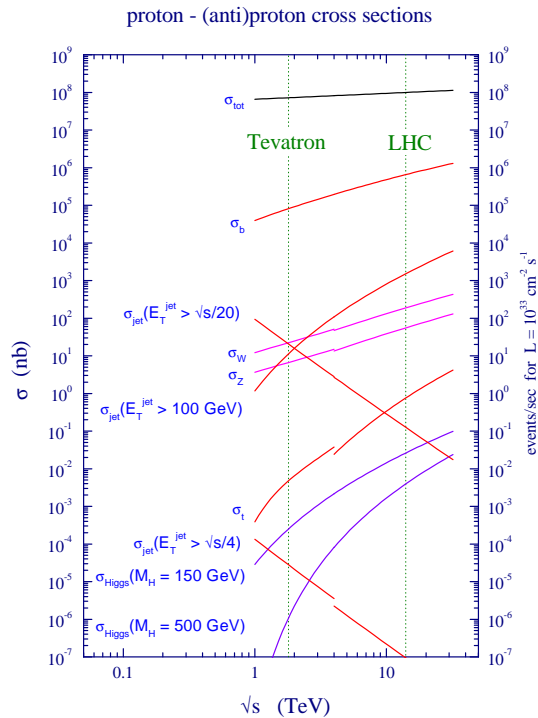}
\vspace*{-1mm}
\caption{ \small
\label{pic:xsect}
Predictions for hard-scattering cross sections 
in $p{\bar p}$ collision at Tevatron and in $pp$ collision at LHC 
as a function of the center-of-mass energy (W.J.~Stirling in Ref.~\cite{Campbell:2006wx}).
}
\vspace*{2mm}
  \end{center}
\end{figure}
The expected cross sections for proton-proton scattering at LHC is large (see Fig.~\ref{pic:xsect}). 
In particular we will have large rates for many Standard Model processes 
such as the production of $b$-quarks, $W^\pm$ and $Z$-bosons, 
jets (even with high cuts on the transverse momentum) and top quarks. 
Much of the physics is actually dominated by the gauge theory of the strong interactions,
Quantum Chromodynamics (QCD). 
Comparing the rates for various processes at Tevatron and LHC in Fig.~\ref{pic:xsect}, 
it is obvious that any search for physics beyond the
Standard Model (BSM) like for superpartners in supersymmetric extensions 
(squarks, gluinos, ...), for Kaluza-Klein modes in models
with extra dimensions or even the search for the Higgs boson 
needs a very precise understanding of the known background from the Standard Model.
Thus, new physics searches require precision predictions, most importantly in QCD.
Moreover, the era of LHC implies a change of paradigm. 
We no longer test QCD, rather we use perturbative
QCD as an essential and established part of our theory toolkit.

Hard QCD is a large subject and, necessarily, the coverage here has to selective 
(see~\cite{Campbell:2006wx,Dixon:2007hh} for other recent reviews on this topic).
In this article, we will briefly review the physics concepts and theoretical 
framework for hard scattering reactions at hadron colliders, focusing on QCD.
We point out achievements of the past years as well as open problems.
We will briefly explain the property of factorization and discuss the parton
luminosity in proton collisions. 
We will summarize the present knowledge on hard parton scattering cross
sections for the production of $W^\pm$ and $Z$ gauge bosons, 
jets in QCD, heavy quarks, like top and bottom and the Higgs.
We put particular emphasis on exact calculations of radiative corrections in QCD 
to next-to-leading order (NLO), next-to-next-to-leading order (NNLO) or beyond. 
Many other aspects such as e.g. the Monte Carlo approach to 
modeling hadronic interactions and parton showers we can only touch briefly 
or else, have to refer to the literature.

With many Standard Model processes to be measured in the early days of LHC 
and the associated uncertainties to be understood on the way to the discovery
of new physics~\cite{Gianotti:2005fm}, we hope that this review serves 
to illustrate a few aspects common to the underlying QCD dynamics.

%
% -----------------------------------------------------------------------------
%
\section{Perturbative QCD at colliders}
\label{sec:pQCD}
The basic prerequisite for the application of perturbative QCD at colliders
is factorization for hard scattering processes.
For hard hadron-hadron scattering this property implies that the 
constituent partons from each incoming hadron interact at short distance 
(i.e. at large momentum transfer $Q^2$).
The property of QCD factorization rests on the fact that we can 
separate the sensitivity to dynamics from different scales (see e.g.~\cite{Collins:1989gx}).
Thus, for a cross section $\sigma_{pp \to X}$ of some 
hadronic final state $X$ in, say, proton-proton scattering we can write
\begin{eqnarray}
\label{eq:QCDfactorization}
  \displaystyle
  \sigma_{pp \to X} = 
  \sum\limits_{ijk}\,
  \int
  dx_1\, dx_2\, dz\,
  f_{i}(x_1,\mu^2) \,
  f_{j}(x_2,\mu^2) \,
  \hat{\sigma}_{ij \to k} \left(x_1,x_2,z,Q^2,\alpha_s(\mu^2),\mu^2 \right)\, 
  D_{k \to X}(z,\mu^2)
  \, ,
\end{eqnarray}
where all functions have a clear physical interpretation. 

\begin{figure}[tb]
  \begin{center}
    \includegraphics[width=12.5cm,angle=0]{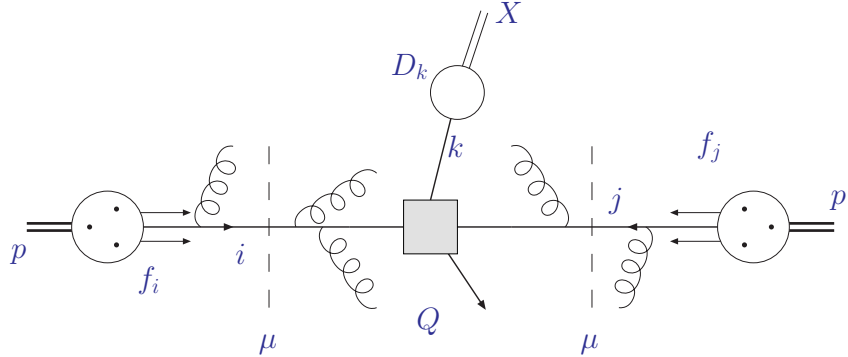}
\vspace*{-1mm}
\caption{ \small
\label{pic:qcd-fact}
Factorization for the hard-scattering cross sections in
Eq.~(\ref{eq:QCDfactorization}) in the QCD improved parton model.
}
\vspace*{2mm}
  \end{center}
\end{figure}

The parton distribution functions (PDFs) in the proton $f_{i}$ ($i=q,{\bar q},g$) 
describe the fraction $x_i$ of the hadron momentum carried by the quark or
gluon and the convolution of $f_{i}$ and $f_{j}$ determines the parton
luminosity at the collider.
The PDFs cannot be calculated in perturbation theory due to the proton being 
a very complicated multi-particle bound state. 
Rather, they have to be obtained from global fits to experimental data.
The (hard) parton cross section $\hat{\sigma}_{ij \to k}$ depending on the parton types $i$, $j$ and $k$ 
is calculable perturbatively in QCD in powers of the strong coupling constant $\alpha_s$ 
and describes how the constituent partons from incoming protons interact at short distances 
of order ${\cal O}(1/Q)$.
The final state $X$ may denote hadrons, mesons, jets, etc. and 
needs another transition from the perturbative hard partons in the final state 
to the observed particles.
The necessary function $D_{k \to X}$ can therefore be a fragmentation function 
or also a jet algorithm. 
Here the interface with showering algorithms (based on a Monte
Carlo approach) becomes particularly crucial.
All quantities in Eq.~(\ref{eq:QCDfactorization}) depend on the renormalization and factorization scale, 
$\mu_r$ and $\mu_f$, which are usually taken to be the same. 
Throughout this review we set $\mu_r = \mu_f = \mu$. 
The details of the integration range in the convolution 
in Eq.~(\ref{eq:QCDfactorization}) are controlled by the kinematics of the hard scattering process.
Schematically QCD factorization can be depicted as in Fig.~\ref{pic:qcd-fact}.

Physical observables like the cross section $\sigma_{pp \to X}$ in Eq.~(\ref{eq:QCDfactorization}) 
cannot depend the factorization scale.
In the perturbative approach, this implies that any dependence on $\mu$ in $\sigma_{pp \to X}$ 
has to vanish at least to the order in $\alpha_s$ considered.
This property can be cast in the following form,
\begin{eqnarray}
\label{eq:fact-scale-dep}
  \displaystyle
  \frac{d}{d \ln\mu^2} \sigma_{pp \to X} = {\cal O}(\alpha_s^{l+1})
  \, .
\end{eqnarray}
It defines the commonly adopted approach to quantify uncertainties in
theoretical predictions based on the variation of the renormalization and factorization scale. 

Let us briefly turn to hard scattering cross sections.
There exist various approaches to the calculation of $\sigma_{pp \to X}$ 
ranging from easy to difficult as far as computational
complexity is concerned as well as from inclusive to fully differential in
terms of kinematical variables.
First of all, there exist parton shower Monte Carlos
(e.g. Herwig~\cite{Corcella:2002jc}, Pythia~\cite{Sjostrand:2006za,Sjostrand:2007gs}, Sherpa~\cite{Gleisberg:2003xi}) 
which are very important tools for understanding multi-parton scattering and the underlying event.

For predictions building on exact matrix elements at leading order (LO), 
we have at our disposal many automated tree level calculations in the Standard Model, 
in its minimal supersymmetric extension (MSSM) or in other
BSM models utilizing programs like e.g. Alpgen~\cite{Mangano:2002ea}, CompHEP~\cite{Boos:2004kh}, 
Helac-Phegas~\cite{Cafarella:2007pc}, MadGraph~\cite{Maltoni:2002qb} or Whizard~\cite{Kilian:2007gr}.
These tools provide first estimates for hard scattering cross sections  
through numerical phase space integration of the exact matrix elements.
In this way, they are flexible as far as kinematics and the topology 
of a given hard scattering observable is concerned 
and allow easy interfacing of LO calculations with parton shower 
Monte Carlos and, possibly, detector simulation. 
However, scattering reactions with exact matrix elements 
for more than 8 jets (particles) in the final state 
are currently at the edge of computational capabilites.

At NLO level we do have some analytical (or numerical) calculations of Feynman
diagrams yielding parton level Monte Carlos (e.g. NLOJET++~\cite{Nagy:2001fj,Nagy:2003tz} or MCFM~\cite{Campbell:2000bg}).
However, we have also seen recently significant progress based on string inspired techniques.
At the edge of technical developments is the concept of exact NLO calculations
interfaced with parton shower in programs as realized in MC@NLO~\cite{Frixione:2003ei,Frixione:2006gn},
POWHEG~\cite{Nason:2004rx,Frixione:2007vw} or VINCIA~\cite{Giele:2007di}. 

At higher orders in QCD perturbation theory, like NNLO some selected results
are known mostly for inclusive kinematics. 
However, in view of LHC, we have witnessed significant progress in the last years 
to provide also predictions in completely differential kinematics~\cite{Anastasiou:2005qj,Catani:2007vq}.
Beyond this level of accuracy at, say, next-to-next-to-next-to-leading order 
(N$^3$LO) only very few results are known, e.g. for deep-inelastic scattering (DIS)~\cite{Vermaseren:2005qc}.

%
% -----------------------------------------------------------------------------
%
\section{Parton luminosity at hadron colliders}
\label{sec:pdf-lumi}

The parton luminosity in Eq.~(\ref{eq:QCDfactorization})
is an indispensable ingredient of hard-scattering processes involving initial-state hadrons.
At hadron colliders one has wide-band beams of quarks and gluons and, as is well known,  
the necessary PDFs of the proton $f_{i}$ ($i=q,{\bar q},g$) 
are not directly accessible in QCD perturbation theory. 
However, the scale dependence (evolution) of PDFs is governed 
by the splitting functions and predicted in a perturbative expansion in powers of $\alpha_s$.
The universality allows for the determination of sets of PDFs in global fits to experimental data.
Upon evolution this information from fits to reference processes 
can be used to provide cross section predictions at LHC energies
and we can quantify the present uncertainties.

\subsection{Parton evolution}
The parton distributions in the hadron are distinguished by the flavor quantum numbers,
which are additive. The valence distribution originates from differences of
quarks and anti-quarks $q - \bar q$.
The proton is composed of the sea distribution (i.e. the sum over all flavors $q + \bar q$) 
and of the gluon $g$.

\begin{figure}[tb]
  \begin{center}
    \includegraphics[width=10.0cm,angle=0]{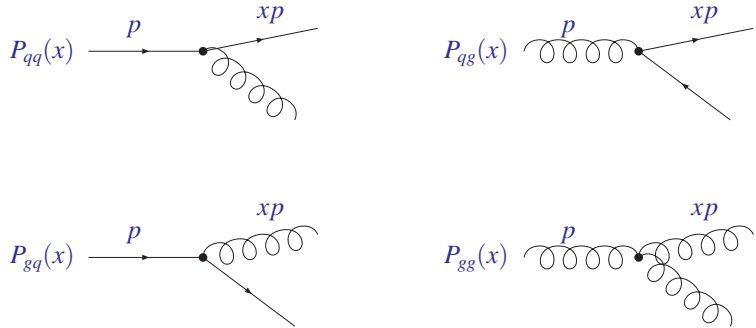}
\vspace*{-1mm}
\caption{ \small
\label{pic:P-diags}
Sample of Feynman diagrams for parton-parton splitting in leading order QCD.
We indicate the collinear momentum flow ($p$ incoming and $xp$ outgoing) 
as it enters the calculation of the corresponding splitting function $P_{ij}$. 
See e.g. Ref.~\cite{Collins:1981uw} for an operator definition of parton distributions.
}
\vspace*{2mm}
  \end{center}
\end{figure}

\begin{table}
  \begin{center}
\begin{tabular}{l c c c c}
\hline\\[-2ex]
 &tree & 1-loop & 2-loop & 3-loop \\
\hline\\[-2ex] 
$q \gamma$  & 1 &  3 &  25 &  359 \\
$g \gamma$  &   &  2 &  17 &  345 \\
$q W$       & 1 &  3 &  32 &  589 \\
$q \phi$    &   &  1 &  23 &  696 \\
$g \phi$    & 1 &  8 & 218 & 6378 \\
\hline\\[-2ex]
sum        & 3 & 17 & 315 & 8367 \\
\hline\\
\end{tabular}
    \caption{ \small
      \label{tab:threeloopP}
      The number of Feynman diagrams contributing to 
      parton ($q,g$)-boson DIS (vector bosons $\gamma/W$ or scalar $\phi$) up to
      three loops.
      The NNLO splitting functions $P_{ij}$ have been 
      determined from the collinear singularity of these scattering
      reactions (see~\cite{Moch:2004pa,Vogt:2004mw}).
    }
  \end{center}
\end{table}
The independence of any physical observable on the scale $\mu$ immediately
gives rise to evolution equations for the PDFs $f_{i}$, $i=q,{\bar q},g$. 
From Eq.~(\ref{eq:fact-scale-dep}) we find that the scale dependence of $f_{i}$ is governed by 
\begin{equation}
\label{eq:evolution}
\frac{d}{d \ln \mu^2}\, 
  \left( \begin{array}{c} f_{q_i}(x,\mu^2)  \\ f_{g}(x,\mu^2)  \end{array} \right)
\: =\: 
\sum_{j}\, \int\limits_x^1\, {dz \over z}\,
\left( \begin{array}{cc} P_{q_iq_j}(z) & P_{q_ig}(z) \\
  P_{gq_j}(z) & P_{gg}(z) \end{array} \right)\,
  \left( \begin{array}{c} f_{q_j}(x/z,\mu^2)  \\ f_{g}(x/z,\mu^2)  \end{array} \right)
\, ,
\end{equation}
which is a system of coupled integro-differential equations 
corresponding to the different possible parton splittings, 
see e.g. Fig.~\ref{pic:P-diags} where some Feynman diagrams contributing 
in leading order QCD are displayed.
The splitting functions $P_{ij}$, i.e. the kernels of these differential equations 
are universal quantities and can be calculated in perturbation theory 
from the collinear singularity of any hard scattering process. 
Thus, $P$ has an expansion in powers of $\alpha_s$ as 
\begin{eqnarray}
\label{eq:P-exp-alphas}
  P & = & 
  \alpha_s\, P^{(0)} + \alpha_s^2\, P^{(1)} + \alpha_s^3\, P^{(2)} + \, \ldots 
  \, ,
\end{eqnarray}
where we have suppressed parton indices.
The first two terms are needed for NLO predictions, 
which is the standard approximation, although often still with large uncertainties.
Currently, the splitting functions are known to NNLO 
and in Tab.~\ref{tab:threeloopP} we give the number of Feynman diagrams for 
the corresponding hard parton reactions in DIS from which the NNLO expressions
$P^{(2)}$ in Eq.~(\ref{eq:P-exp-alphas}) have been calculated~\cite{Moch:2004pa,Vogt:2004mw}.

\begin{figure}[htb]
  \begin{center}
  \includegraphics[bbllx=50pt,bblly=285pt,bburx=355pt,bbury=500pt,angle=0,width=7.5cm]{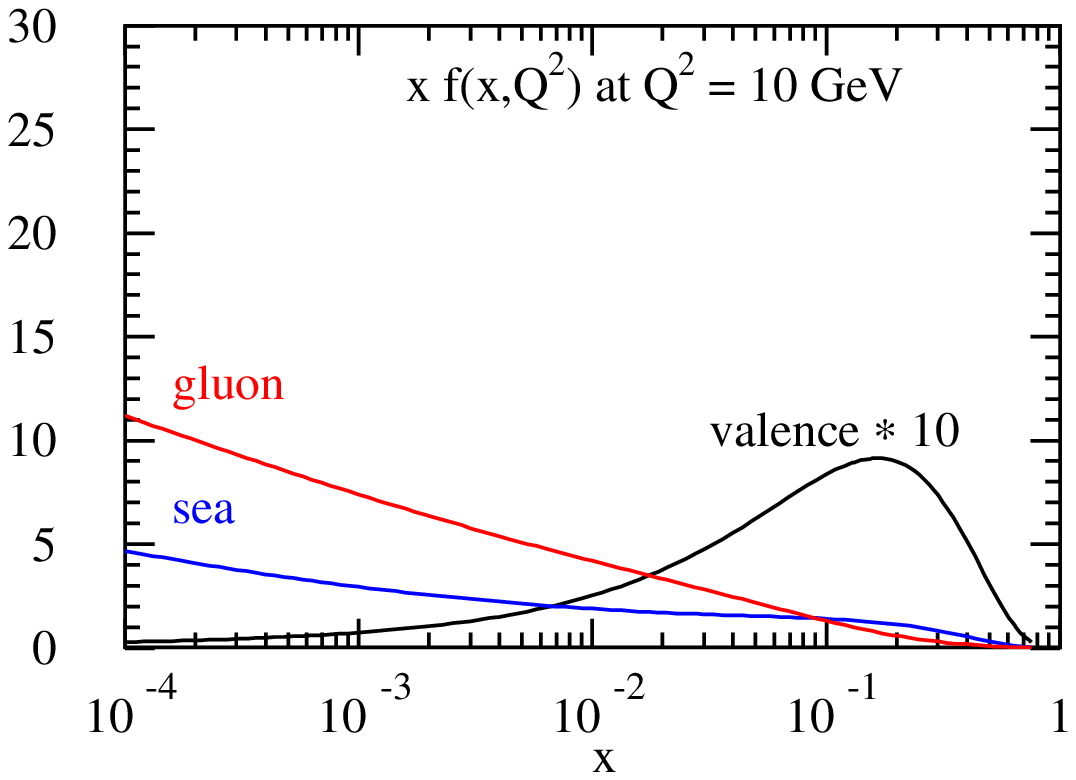}
\hspace*{5mm}
  \includegraphics[bbllx=50pt,bblly=285pt,bburx=355pt,bbury=500pt,angle=0,width=7.5cm]{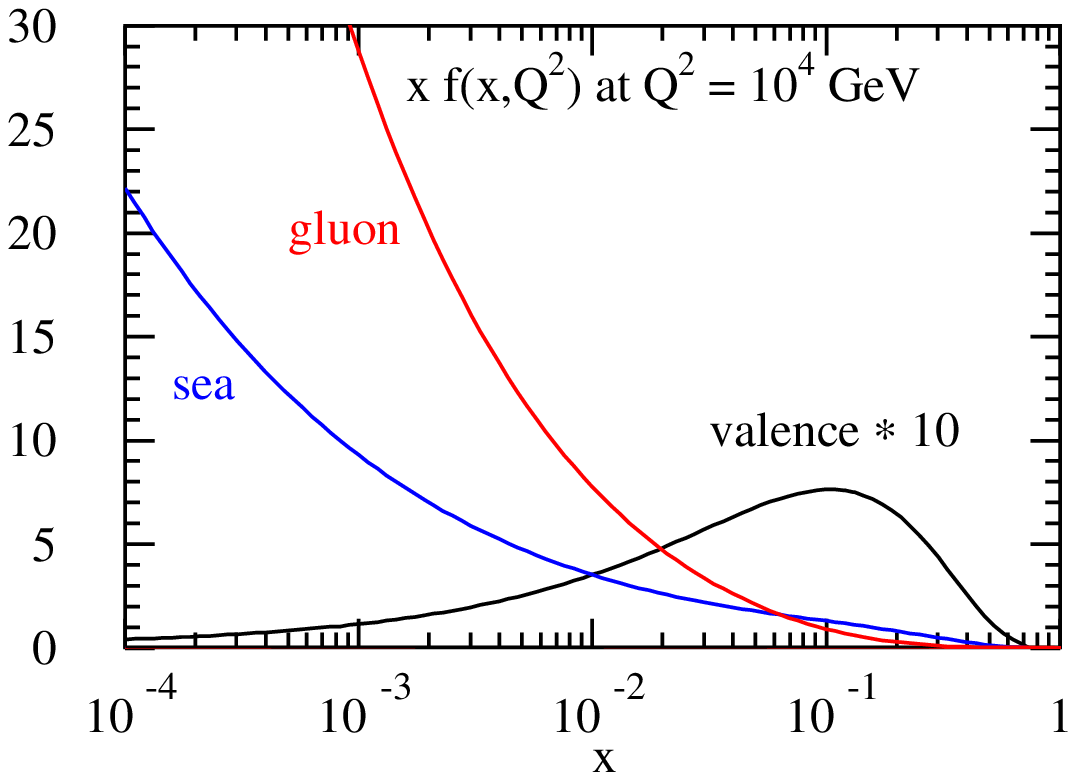}
\vspace*{-1mm}
\caption{ \small
\label{pic:pdf}
Evolution of the valence, sea and gluon momentum distributions $x f(x,Q^2)$ in the proton 
from a low scale at $Q^2 = 10$~GeV (left) to LHC energies at $Q^2 = 10^4$~GeV (right) 
for the parameterization of Ref.~\cite{Martin:2004ir} showing the strong rise
of the gluon at small $x$.
}
\vspace*{2mm}
  \end{center}
\end{figure}

Physically, the evolution Eq.~(\ref{eq:evolution}) states that one becomes sensitive to lower momentum partons
as the resolution of the proton is increased, i.e. as the scale $\mu$ becomes larger.
Given an input distribution at a low scale, say $Q^2 = 10$~Gev$^2$, 
which has to be determined in a global fit from comparison to data,
one can solve Eq.~(\ref{eq:evolution}) to predict the PDFs at a high scale (see Fig.~\ref{pic:pdf}).
Solutions of Eq.~(\ref{eq:evolution}) can be obtained by a variety of 
methods with available codes~\cite{Vogt:2004ns,salam:hoppet,botje:qcdnum} 
and benchmarks are provided in Refs.~\cite{Giele:2002hx,Dittmar:2005ed}.

Modern parameterizations of parton distribution from global fits account in particular 
for the effects of experimental errors and come with the according uncertainties, 
see e.g. the framework LHAPDF accord~\cite{Giele:2002hx,Whalley:2005nh,cedar:lhapdf}.
Much of the needed experimental information originates from deep-inelastic scattering 
data on structure functions from HERA for $e^\pm p$-scattering (H1, ZEUS) and 
from fixed targets (proton and deuterium) for $\mu p$ and $\mu d$ scattering (BCDMS, NMC, SLAC, E665),
as well as (anti-)neutrino-proton scattering (CCFR), see e.g.~\cite{Yao:2006px}.
These data determine the quark distributions for light flavors at all momentum fractions $x$ and
through the scale evolution in perturbative QCD the gluon distribution at medium and small $x$.
Further information on the flavor content of the nucleon is provided by
structure function data for $F_2^{charm}$ from HERA for the charm distribution
and by Drell-Yan data on proton-nucleon targets (E605, E772, E866), 
which determine the sea quark distributions, in particular $f_{\bar u}$ and $f_{\bar d}$.
The Tevatron experiments CDF and D0 are able to constrain the ratio $f_u/f_d$ at high $x$ 
with the rapidity asymmetry in $W$-boson production and 
the gluon distribution at high $x$ with the help of inclusive jet data.
More recently, by relaxing the assumption $f_{s}=f_{\bar s}$, 
also information on the strange asymmetry $f_{s}$ and $f_{\bar s}$ has
been extracted from $\nu({\bar \nu}) p$ scattering (NuTeV, CCFR).

\subsection{Parton distributions from HERA to LHC}
Given the great importance of deep-inelastic scattering data a question that
has been frequently addressed in the past is, of course, the impact HERA data 
for LHC predictions, in particular as far as the parton luminosity is concerned
(see Refs.~\cite{Dittmar:2005ed,Vogt:2007vv} for further discussion).
\begin{figure}[htb]
  \begin{center}
    \includegraphics[width=8.0cm,angle=0]{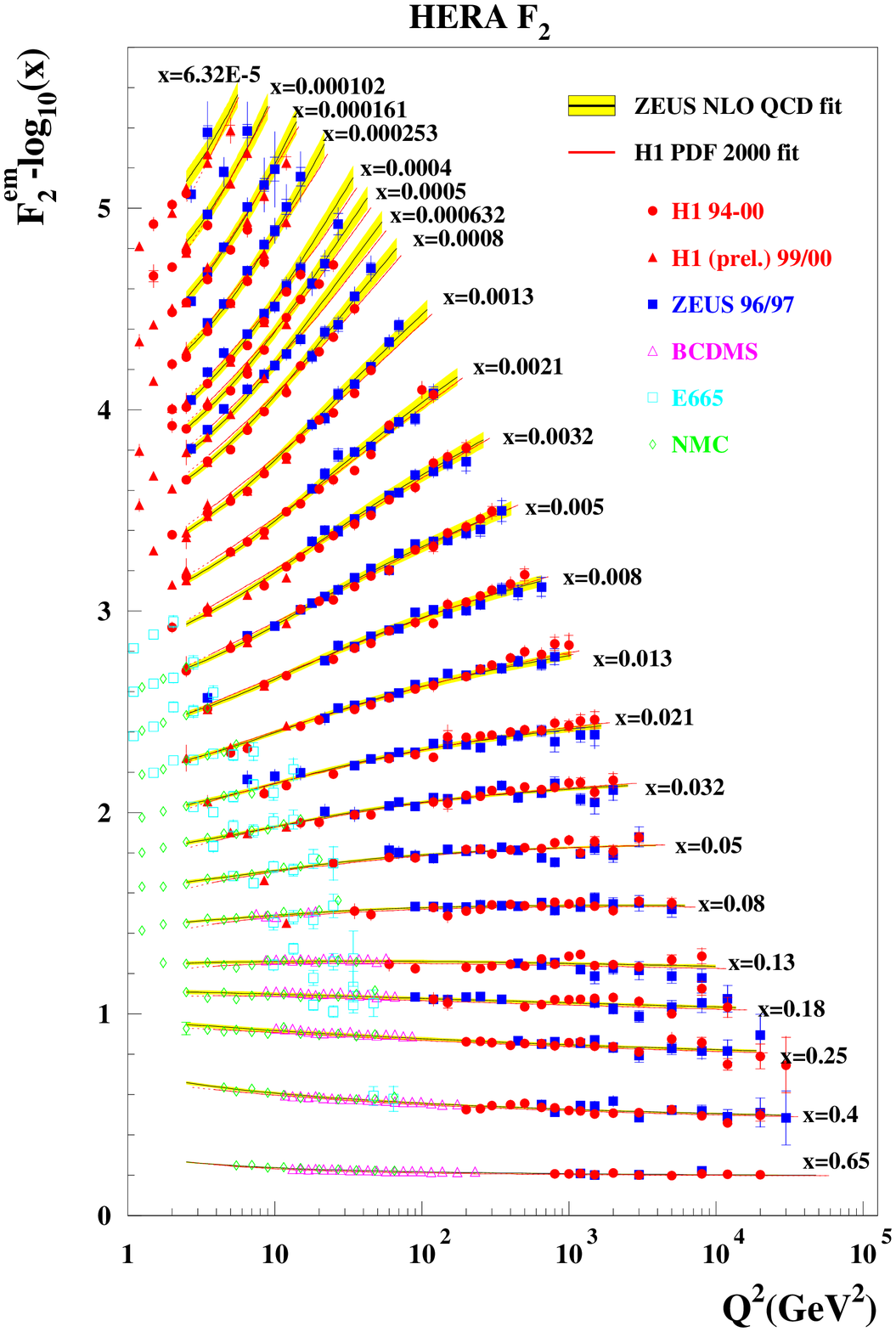}
    \includegraphics[width=8.0cm,angle=0]{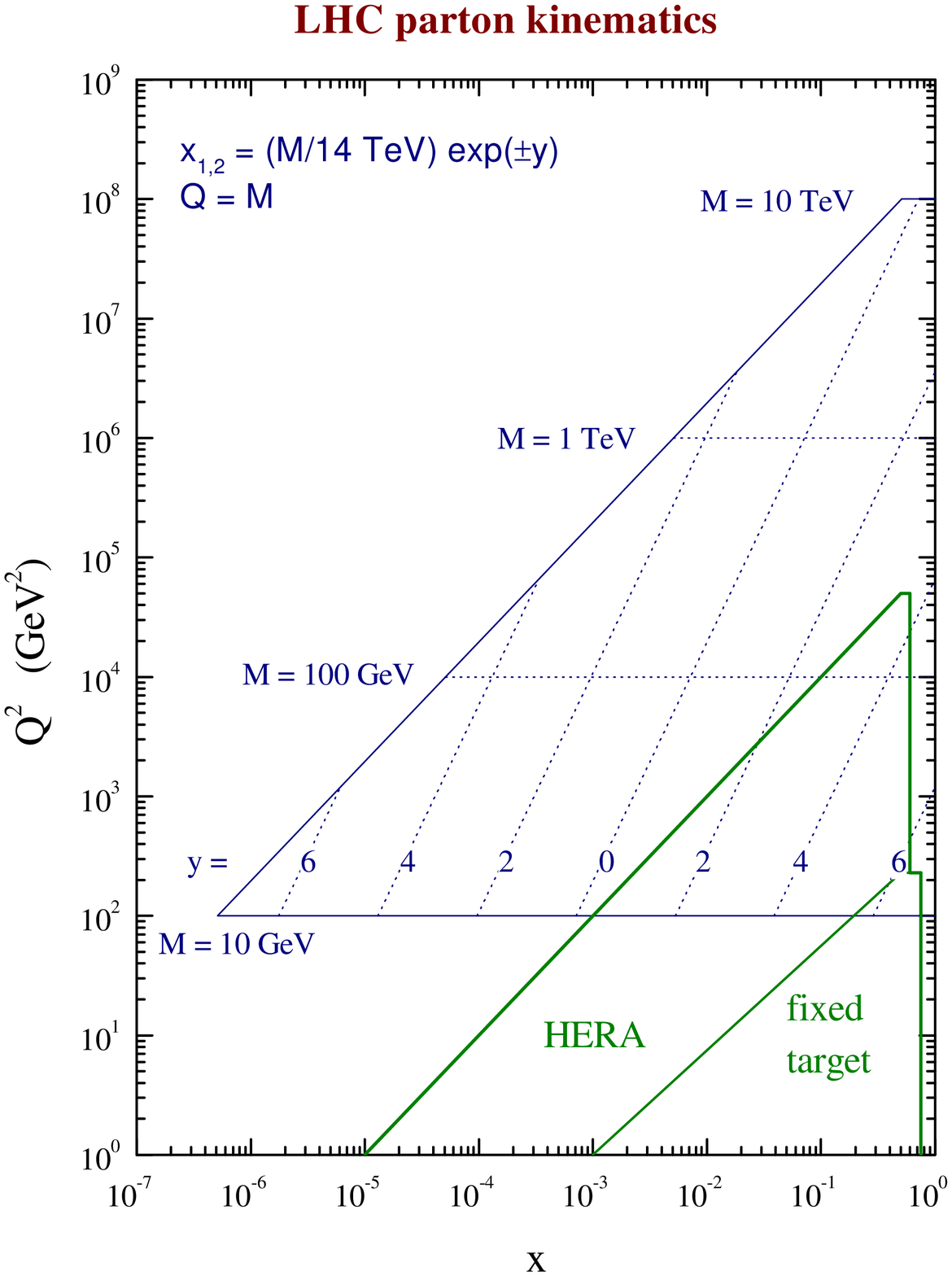}
\vspace*{-1mm}
\caption{ \small
\label{pic:lhckin}
Left: The structure function $F_2$ in deep-inelastic $e^\pm p$ scattering as a function of $x$ and $Q^2$ 
(from Ref.~\cite{Chekanov:2002pv}).
Right: Parton kinematics at LHC and at HERA (W.J.~Stirling in Ref.~\cite{Campbell:2006wx}).
}
\vspace*{2mm}
  \end{center}
\end{figure}

As illustrated by the compilation of measurements in Fig.~\ref{pic:lhckin} (left), 
the data on the structure function $F_2$ in deep-inelastic $e^\pm p$ scattering 
extend over a wide range in $x$ and $Q^2$. 
Considering, on the other hand, the allowed region for parton kinematics at LHC 
in Fig.~\ref{pic:lhckin} (right), it is clear, that there is a large overlap in $x$ 
with the range covered by HERA.
However the relevant hard scale $Q$ is typically two to three orders higher 
due to the increased center-of-mass energy ${\sqrt S}$.
Details of the parton kinematics at LHC depend, of course, on the invariant mass $M$ of the final state 
and on the rapidity $y$. 
The dominant values of the momentum fractions are $x_{1,2} \sim (M e^{\pm y})/{\sqrt S}$ 
and variation of $M$ and $y$ at fixed ${\sqrt S}$ tests the sensitivity to partons
with different momentum fractions.
\begin{figure}[htb]
  \begin{center}
    \includegraphics[width=12.5cm,angle=0]{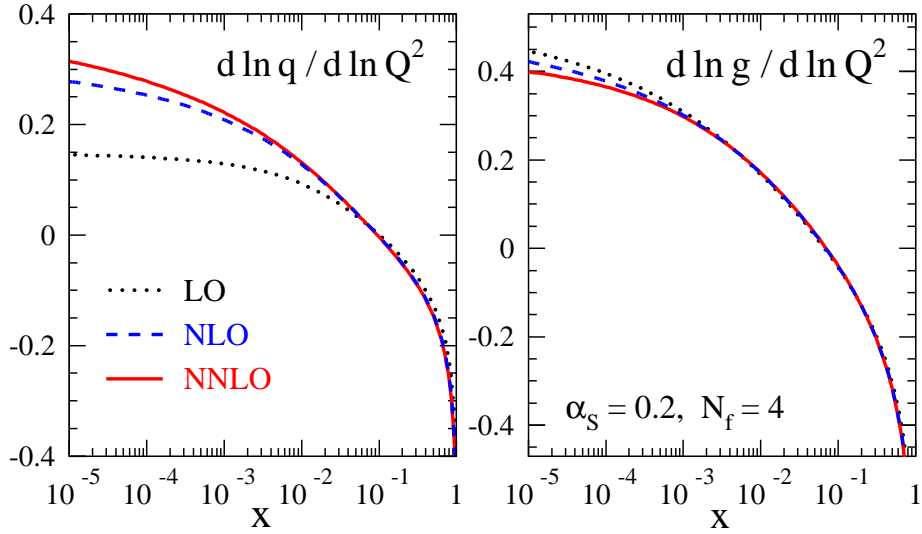}
    \vspace*{-1mm}
\caption{ \small
\label{pic:nnlo-evolution}
Perturbative expansion of the scale derivatives of typical quark
 and gluon distributions at $\mu^2 \approx 30$ GeV$^2$ (from 
 Ref.~\cite{Vogt:2004mw}, where the initial conditions are specified).
}
\vspace*{2mm}
  \end{center}
\end{figure}

The large difference in the hard momentum scale $Q$ between HERA and LHC requires 
the parton evolution based on Eq.~(\ref{eq:evolution}) to be sufficiently accurate 
in perturbative QCD.
The necessary perturbative accuracy for quantitative predictions is approached 
at NNLO~\cite{Moch:2004pa,Vogt:2004mw}. 
The stability of evolution is shown in Fig.~\ref{pic:nnlo-evolution}, 
where the scale derivatives of quark and gluon distributions at $\mu^2 \approx 30$ GeV$^2$ 
are displayed. 
Obviously, the expansion is very stable except for very small momentum fractions 
$\:x \lsim 10^{-4}$ which shows that the perturbative evolution Eq.~(\ref{eq:evolution}) 
is applicable down to very small $x$.
In terms of LHC parton kinematics, this corresponds to perturbative stability 
for central rapidities $|y| \lsim 2$, 
while modifications are at most expected in the very forward (backward) regions $|y| \gsim 4$.

\subsection{$W$ and $Z$-boson production at LHC}
\begin{figure}[htb]
  \begin{center}
    \includegraphics[width=8.0cm,angle=0]{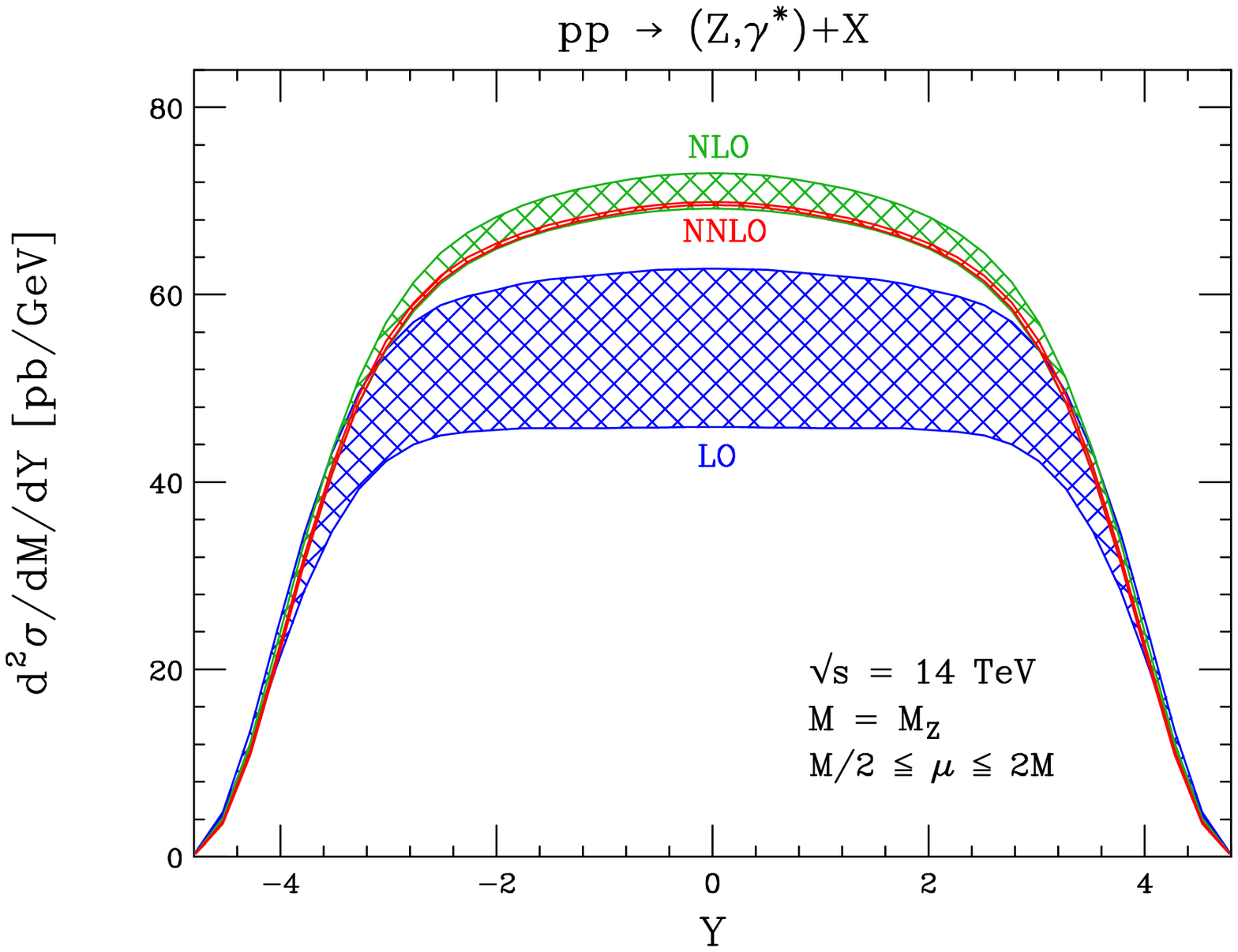}
    \includegraphics[width=8.0cm,angle=0]{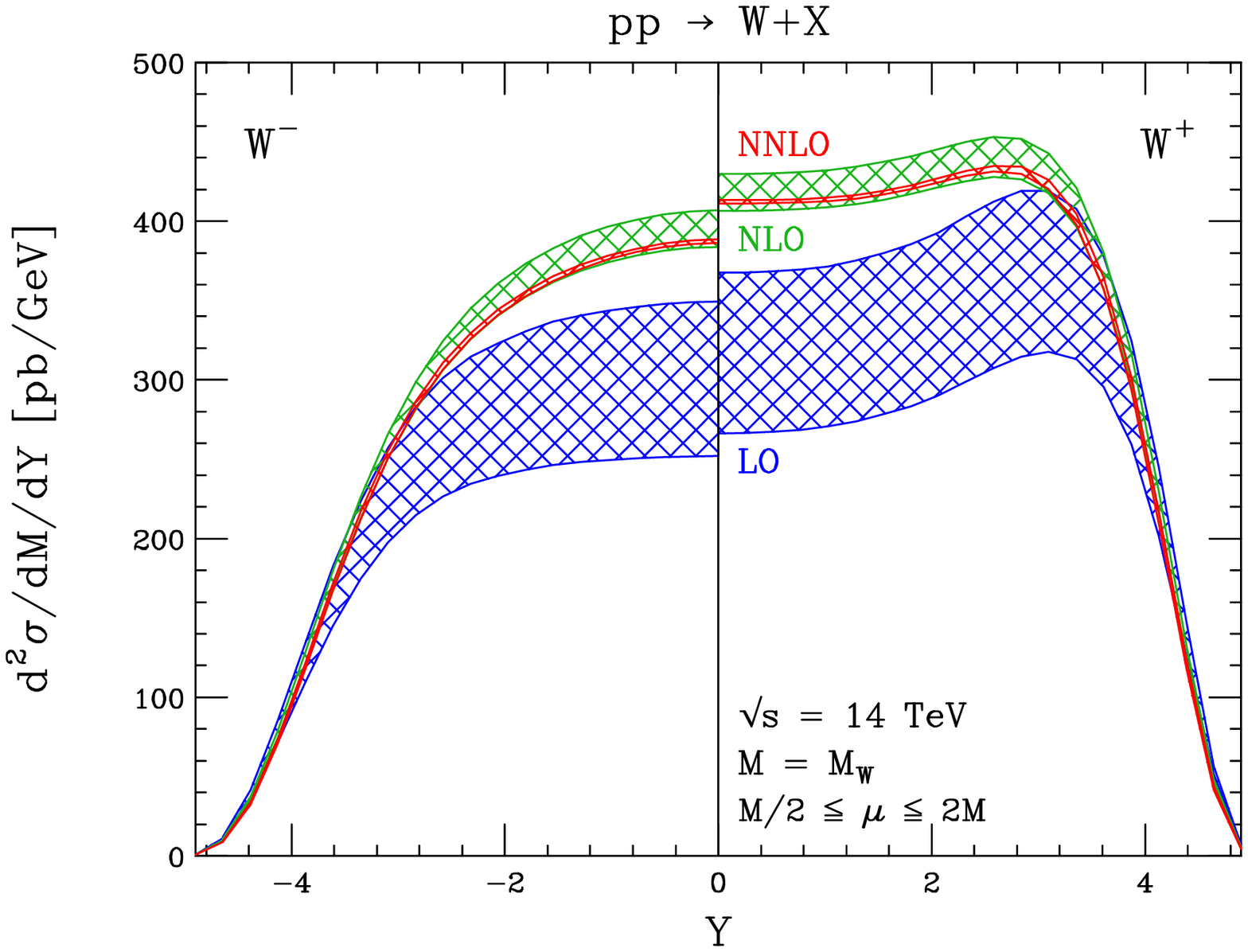}
\vspace*{-1mm}
\caption{ \small
\label{pic:lhc-w-rap}
The rapidity-dependent cross sections for gauge-boson production at the 
LHC, using the partons of Ref.~\cite{Martin:2002dr} 
and estimates of the theoretical uncertainty from variations of the scale $\mu$
(from Ref.~\cite{Anastasiou:2003ds}).
}
\vspace*{2mm}
  \end{center}
\end{figure}

The immediate question arises: What is the impact of our current knowledge of 
parton distributions on the precision of LHC predictions, 
for instance for $W^\pm, Z$-boson rapidity distributions, 
which often have been considered {\it "standard candle"} 
processes for the parton luminosity~\cite{Dittmar:2005ed,Dittmar:1997md}.
The corresponding cross sections are known to NNLO in perturbative QCD, 
and according to Eq.~(\ref{eq:fact-scale-dep}) one can quantify the 
theoretical uncertainties obtained by varying the 
renormalization and factorization scale $\mu$ by the conventional (although arbitrary) 
factor of two around $M_{W,Z}$. 
The perturbative stability of the results in Fig.~\ref{pic:lhc-w-rap} nicely demonstrates 
the necessity of considering higher order perturbative corrections through NNLO in QCD.
It would be impossible to make precision predictions, or perform precision analyses, 
based solely on the rough (and non-overlapping) LO and NLO error estimates.
\begin{figure}[htb]
  \begin{center}
    \includegraphics[width=5.25cm,angle=0]{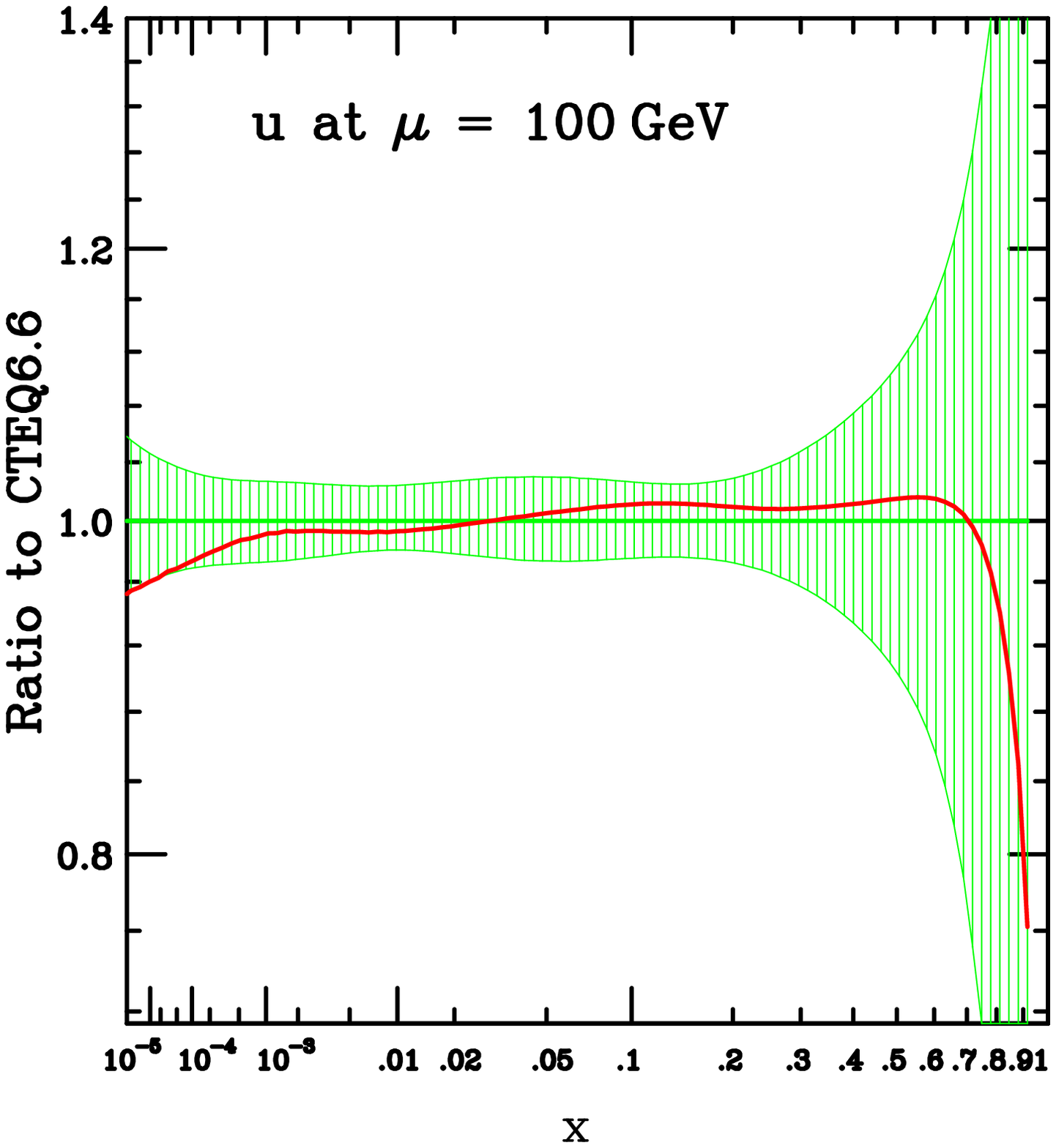}
    \includegraphics[width=5.25cm,angle=0]{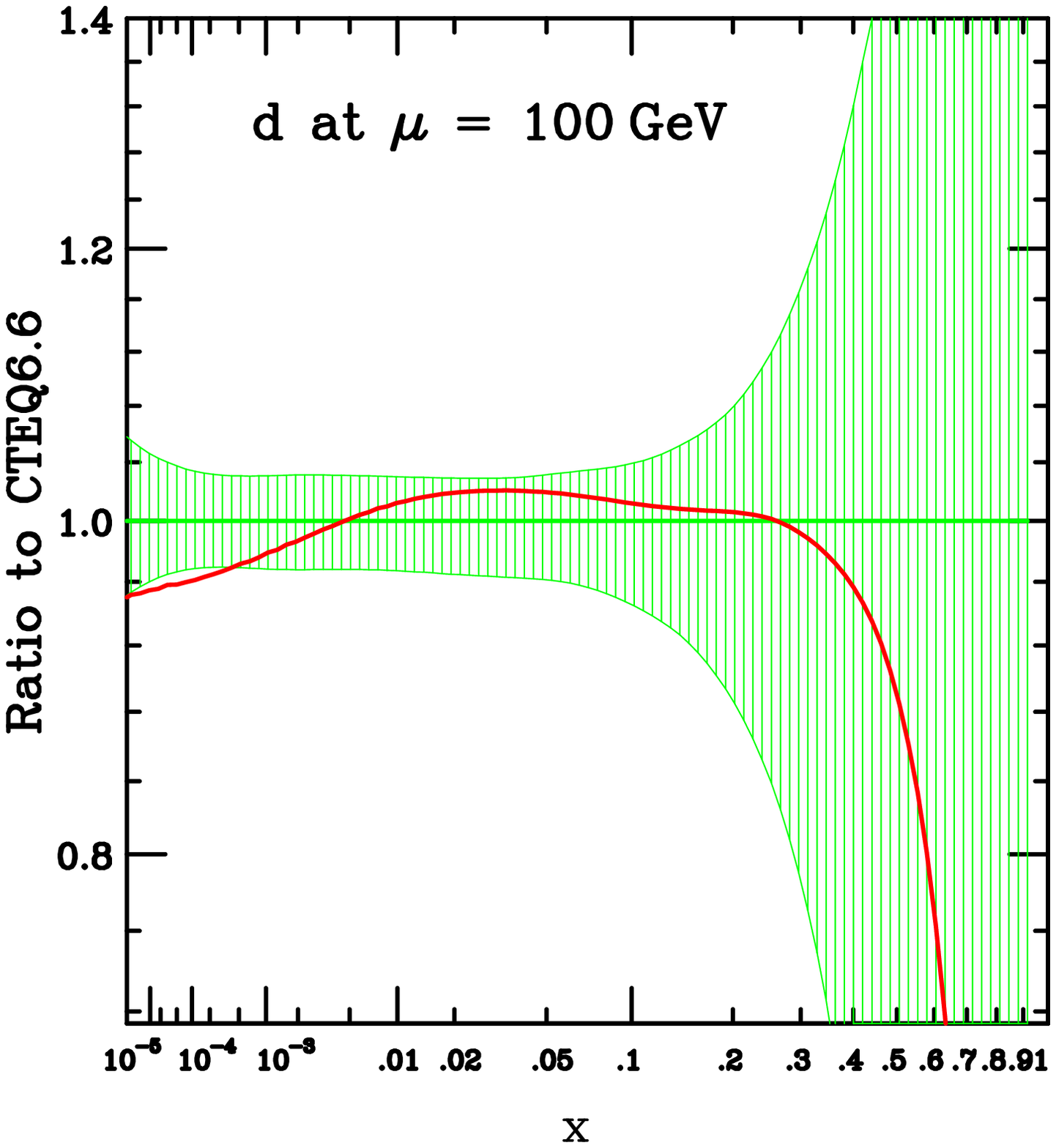}
    \includegraphics[width=5.25cm,angle=0]{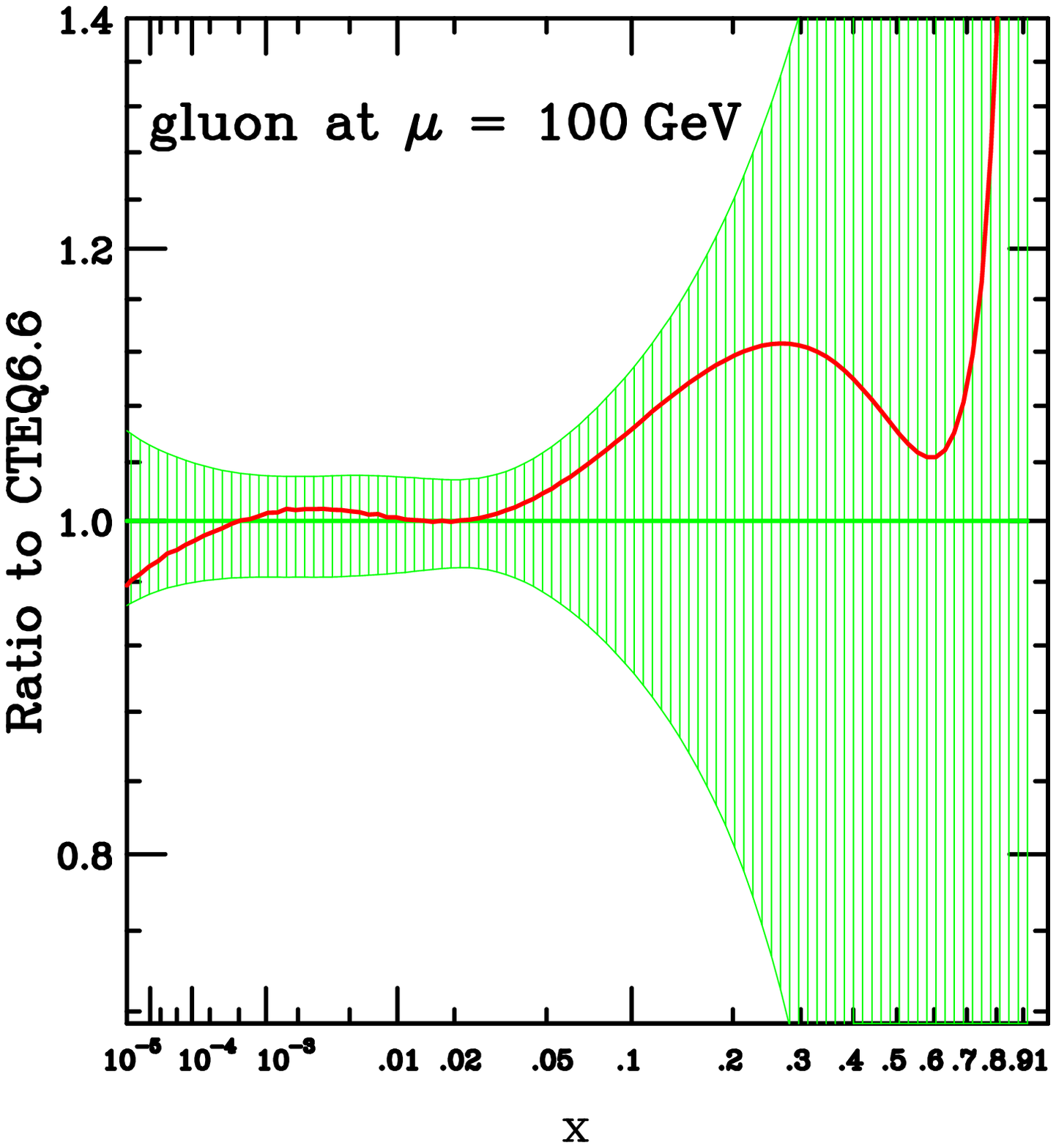}
\vspace*{-1mm}
\caption{ \small
\label{pic:cteq-pdf-changes}
The distributions $f_u$, $f_d$ and $f_g$ of the CTEQ6.6 fit~\cite{Nadolsky:2008zw}) 
with the estimated error bands (shaded area) at the scale $\mu=100$ GeV. 
Comparison with previous results (CTEQ6.1M)~\cite{Pumplin:2002vw} is denoted
by solid lines (from Ref.~\cite{Nadolsky:2008zw}).
}
\vspace*{2mm}
  \end{center}
\end{figure}
Recent improvements in the parameterizations of PDFs, though, have been shown
to significantly affect predictions for physical cross sections at LHC.
An independent treatment of the strange quark distributions $f_{s}$ and $f_{\bar s}$ 
(hence their uncertainties), for instance, has an impact on the correlated uncertainties of the light sea quarks, 
because neutral current deep-inelastic data on $F_2$ constrains the combination 
$4/9 (f_{u}+f_{\bar{u}})+1/9(f_{d}+f_{\bar{d}}+f_{s}+f_{\bar{s}})$.
In consequence, the size of the uncertainty on the sea quarks for values $x\sim 10^{-3}-10^{-2}$ 
at hard scales $Q^2\sim M_W^2$ roughly doubles from $\sim 1.5\%$ to $\sim 3\%$ 
for the MSTW group~\cite{Thorne:2007bt,Martin:2007bv}.
CTEQ in their newer sets (e.g. CTEQ6.6) has improved the treatment of the charm contribution to the 
deep-inelastic structure function $F_2$ at HERA by implementing now a general-mass formalism 
for a variable flavor number scheme consistent with QCD factorization, see~\cite{Chuvakin:1999nx,Tung:2001mv}.
The reduced charm component of $F_2$ is compensated by larger light quark distributions
$f_u$ and $f_d$ at small $x$ as illustrated in Fig.~\ref{pic:cteq-pdf-changes}.

As an upshot, the predictions for $W^\pm$- and $Z$-production cross sections at LHC 
being sensitive to PDFs in the $x\sim 10^{-3}$ range shift by $8\%$ 
between the sets CTEQ6.6~\cite{Nadolsky:2008zw} and CTEQ6.1M~\cite{Pumplin:2002vw}.
Although this particular shift originates from theoretical improvements long overdue, 
it is an example that PDFs and their associated uncertainties  
will have a significant impact on the precision of the 'gold-plated' $W^\pm$- and $Z$-cross-sections and 
$W^\pm/Z$-ratio calibration measurement. 
In this context, it should also be stressed, that PDF uncertainties in the
region of very small momentum fractions, $x\simeq 10^{-5}$, (as e.g. displayed
in Fig.~\ref{pic:cteq-pdf-changes}) largely rely upon extrapolations of data 
and represent a certain parameterization bias.

\subsection{Parton distributions and the search for new physics}
\label{pdfs-and-bsm}
Apart from gauge boson production, there are prominent measurements at LHC 
which depend on our knowledge of parton distributions and, in turn, might be used to improve it.
High-$E_t$ jet cross-sections, for instance, are a particularly prominent place to look for BSM effects.
The discovery of new physics, such as e.g. jet signals for low mass
strings~\cite{Anchordoqui:2007da}, large extra dimensions or models parameterized in terms of contact interactions
becomes sensitive to the uncertainty of the gluon PDF especially at low-$x$.
Recently, also top-pair-production has been proposed as an additional calibration process at LHC, 
because its PDF dependence is anti-correlated with $Z$-boson production~\cite{Nadolsky:2008zw} 
and correlated with Higgs boson production, especially for larger Higgs masses.
Presently, however, the sizable theoretical uncertainties at NLO in QCD are
limiting the applicability of this proposal.

\begin{figure}[hbt]
  \begin{center}
    \includegraphics[width=15.0cm,angle=0]{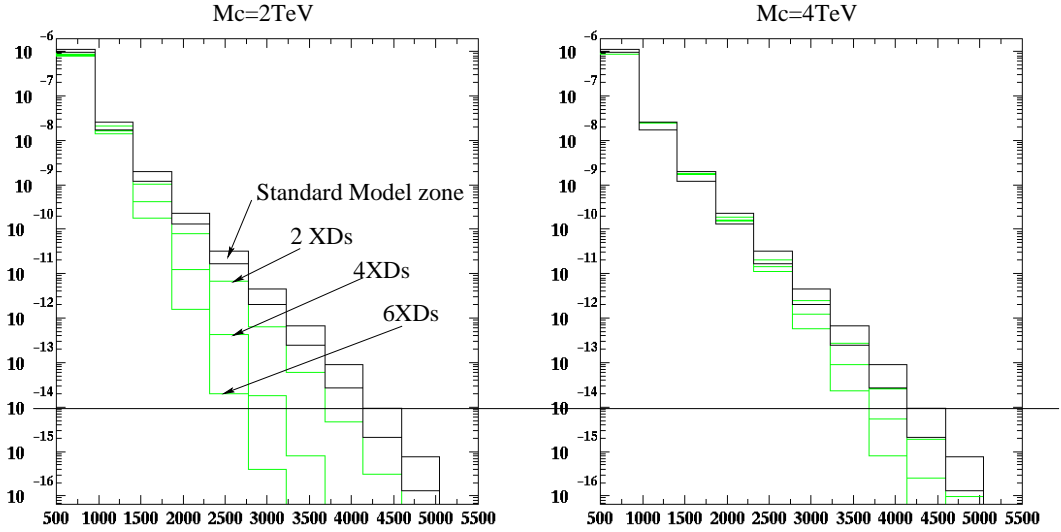}
\vspace*{-1mm}
\caption{ \small
\label{pic:xd-sens-pdfs}
Di-jets cross sections at LHC as a function of $p_t$ for two 
compactification scales $M_c=2$~TeV (left) and $M_c=4$~TeV (right) 
of the extra-dimensions.
Shown are different predictions for various numbers of extra dimensions,
i.e. 2,4 and 6, and the Standard Model zone incorporates the PDF uncertainties of the CTEQ6.1M set.
The horizontal line shows the sensitivity limit corresponding to an LHC
luminosity of $100$~fb$^{-1}$ (from Ref.~\cite{Ferrag:2004ca}).
}
\vspace*{2mm}
  \end{center}
\end{figure}

For di-jet rates at LHC, e.g. the consequences for predictions from large extra dimensions have 
been analyzed~\cite{Ferrag:2004ca} through a modified renormalization group
equation for the strong coupling $\alpha_s$, where the running of $\alpha_s$ 
accelerates due to power corrections as the compactification scale $M_c$ 
of the large extra dimensions is approached. 
Results of a study for the di-jet transverse momentum distribution ($p_t$)
with the event generator Pythia are displayed in Fig.~\ref{pic:xd-sens-pdfs}.
The plot clearly indicates the reduced sensitivity to
extra-dimensions because of the PDF uncertainties in the di-jet $p_t$-distribution. 
Hence there is need to either look at ratios of rates, 
$\sigma_{pp \to 3~\mbox{jets}}$ over $\sigma_{pp \to 2~\mbox{jets}}$ 
or, else at angular correlations of di-jets to reduce the parton luminosity dependence.

On the other hand, PDF uncertainties most likely do not affect the discovery potential of
a Higgs in the mass range $100-1000~{\rm GeV}$ or a high mass $Z^\prime$ in the mass
range $150-2500~{\rm GeV}$. 
Apart form the hadronic di-jets, promising other measurements to be conducted
at LHC itself also include direct photon production to constrain the gluon PDF at low-$x$ 
or the $W^\pm$-asymmetry to obtain information on the low-$x$ valence PDFs.
In particular, the improved description of the (anti-)strange quark distributions leads to 
interesting implications for collider phenomenology. 
For instance the production of a charged Higgs boson $H^{+}$ via the partonic process
$c+{\bar{s}}\to H^{+}$, provides an example of a BSM process 
that is sensitive to the strange PDF in models with two or more Higgs doublets. 
The cross section also depends on a possible intrinsic charm component of the
proton and the recent PDF set CTEQ6.5c provides various models
for such a component~\cite{Pumplin:2007wg}.

%
% -----------------------------------------------------------------------------
%
\section{Parton cross sections}
\label{sec:parton-crs}

Let us next turn to the hard scattering cross sections initiated through the constituent partons $i, j$
of the incoming protons. In general, we consider ${\hat \sigma}_{ij \to X}$ where $X$ denotes any final state allowed
by the Standard Model or its possible extensions. 
Calculations of ${\hat \sigma}_{ij \to X}$ result in predictions for experimental signatures and, eventually, 
determine the power to discriminate BSM signals or, e.g. Higgs boson production, 
from known background of the Standard Model. 
The key issue for QCD theory is the reliability of signal and background estimates.

\subsection{QCD @ NLO}
\label{sec:qcd@nlo}
At a hadron collider, the problem of signal significance has various aspects. 
Experimentally, measurements of hard scattering reactions require reliable identification of leptons (electrons, muons), 
a good understanding hard jets at high transverse momentum ($p_t$), especially
$b$-quark jets, and a sufficiently precise calibration of the jet energy scale.
Moreover, BSM or Higgs searches rely heavily on the presence of large missing
transverse energy ($\slash{E}_t$) to reject Standard Model background compared to the signal.

On the theory side, as briefly mentioned in Sec.~\ref{sec:pQCD}, we have
various levels of accuracy for the hard scattering process 
(assuming that the underlying event and multiple parton interaction are modeled by shower Monte Carlos). 
Estimates to LO in QCD based on exact matrix elements seem mandatory in search scenarios 
for studies of distributions, e.g. in $p_t$ or the (pseudo-)rapidity ($\eta$) and for 
assessing the effects of kinematical cuts.
It is well known, that the overall normalization and, in particular, 
the hard tail of these distributions (e.g at high $p_t$ or $\slash{E}_t$) 
are not well modeled by shower Monte Carlos alone.
However, any LO prediction has large theoretical uncertainties, 
typically estimated by the scale variation, Eq.~(\ref{eq:fact-scale-dep}).
Consider, for instance, the cross section for $pp \to W + 4$~jets, 
which is of ${\cal O}(\alpha_s^4)$ at LO. 
From a variation of the coupling of $\Delta (\alpha_s^{\rm LO}) \simeq 10 \%$
one can roughly estimate a cross section uncertainty of $\Delta (\sigma^{\rm LO}) \simeq 40 \%$.
Thus, one needs to go beyond the Born approximation for certain processes.

Perturbative QCD corrections at NLO to scattering processes are essential for the rates 
and shapes of distributions of Standard Model processes as well as for BSM searches, 
where they may have an impact on the signal significance. 
Often, one encounters large K-factors and also new parton channels open up at NLO 
which may eventually dominate beyond tree level, a prominent example being single-top production 
(see Sec.~\ref{sec:single-top}).
In a series of workshops a number of key processes at LHC has been identified which need to be known
to NLO in QCD. These are summarized in the so-called LHC ``priority'' wishlist in Tab.~\ref{tab:les-houches} 
and the computation of these radiative corrections is presently a very active
field of research.
\begin{table}
  \begin{center}
\begin{tabular}[t]{|l|l|l|}
\hline 
process & background to & reference
\\
($V \in \{ \gamma, W^\pm, Z \}$) & & 
\\
\hline 
$pp \to VV + 1\, \mbox{jet}$ & $t{\bar t} H$, new physics & $WW + 1\,\mbox{jet}$~\cite{Dittmaier:2007th,Campbell:2007ev}
\\
$pp \to H + 2\, \mbox{jets}$ & $H$ production by vector boson fusion (VBF) & $H + 2\,\mbox{jets}$~\cite{Campbell:2006xx}
\\
$pp \to t{\bar t} b{\bar b}$ & $t{\bar t} H$ & 
\\
$pp \to t{\bar t} + 2\, \mbox{jets}$ & $t{\bar t} H$ &
\\
$pp \to VV b{\bar b}$ & $\mbox{VBF}\, \to VV$, $t{\bar t} H$, new physics &
\\
$pp \to VV + 2\, \mbox{jets}$ & $\mbox{VBF}\, \to VV$ & 
\\
$pp \to V + 3\, \mbox{jets}$ & various new physics signatures &
\\  
$pp \to VVV$ & SUSY trilepton & ZZZ~\cite{Lazopoulos:2007ix}, WWZ~\cite{Hankele:2007sb}
\\
%
% Heinrich in Les Houches (2007)
% pp -> b bar b bar
% pp -> gg -> W+ W^-
% NNLO zu VBF, t tbar, Z/gamma + jet, W + jet 
%
\hline 
\end{tabular}
    \caption{ \small
      \label{tab:les-houches}
      Scattering processes at LHC for which the radiative corrections 
      to NLO in QCD are needed, as summarized in Les Houches 2005 (from Ref.~\cite{Buttar:2006zd}).
    }
  \end{center}
\end{table}
\begin{figure}[htb]
  \begin{center}
    \includegraphics[width=15.0cm,angle=0]{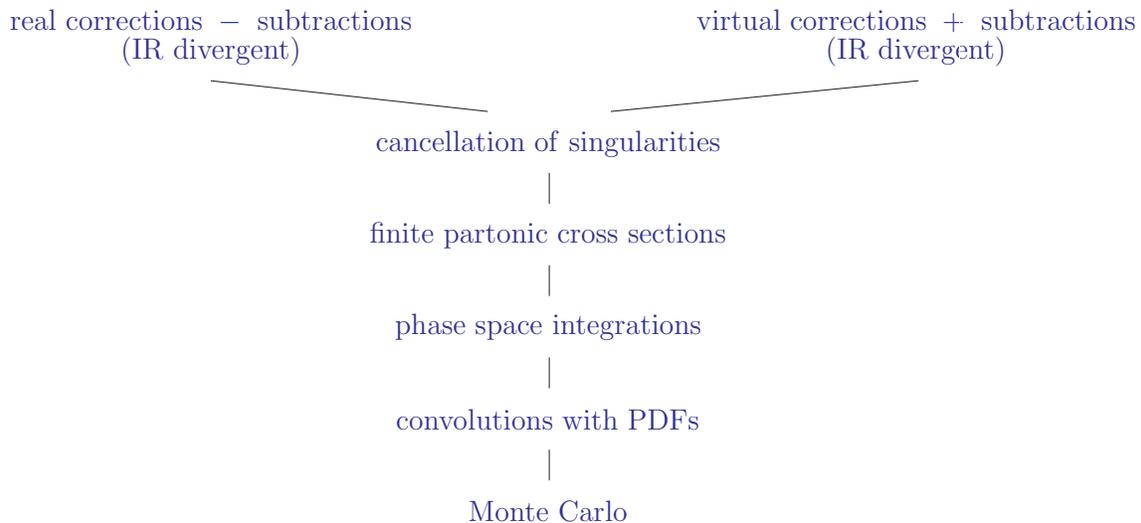}
    \vspace*{-1mm}
    \caption{ \small
      \label{pic:nlo-flowchart}
      Outline of a generic calculation of NLO QCD corrections to a
      (multi-particle) scattering process in the traditional approach.
    }
\vspace*{2mm}
  \end{center}
\end{figure}
The obvious question is, of course, why the calculation of one-loop corrections in QCD is so difficult? 
After all, the conceptual issues are all solved and any computation can follow a straightforward algorithm:
Draw all Feynman diagrams and evaluate them, then use standard reduction techniques for tree and loop amplitudes.
While this is true in principle, it is hard in practice with known bottlenecks, 
because intermediate expressions are much more complicated than the final result.
Thus, let us look at the outline of a generic NLO calculation as displayed in Fig.~\ref{pic:nlo-flowchart}.
For the scattering reaction $2 \to n$~partons, the basic ingredients in
the cross section calculation are the real corrections, 
i.e. the tree level $2 \to (n+1)$~parton reaction $d\sigma^{\rm real}$, 
and the one-loop virtual corrections to the $2 \to n$~parton amplitude, which
are subject to the standard ultraviolet (UV) renormalization.

The latter contribution to the cross section, that is the one-loop virtual correction $d\sigma^{\rm virtual}$, 
in the standard Feynman diagram approach generates large expressions, although we do expect
large cancellations between the diagrams in a gauge theory as a consequence of gauge invariance.
Specifically, one is required to calculate tensor integrals like, e.g.
\begin{eqnarray}
  \label{eq:tensor-int}
  I^{ \mu_1, \mu_2, \dots}(k_1,\dots) \,=\,
  \int\, d^Dp_1\,
  {p_1^{{ \mu_1}} p_2^{{ \mu_2}} { \dots}
    \over (p_1^2-m_1^2) ((p_1-k_1)^2-m_2^2) \, \dots\,  
  }
  \, .
\end{eqnarray}
Unfortunately, these become rather complicated for five or more external
particles, one well-known problem being the numerical stability for all allowed 
configurations of the external momenta.
Without going into details here, suffice it to say, that reduction algorithms for tensor 
integrals are some 30 years after the work of Passarino and Veltman~\cite{Passarino:1979jh} still a very active
field of research, see e.g.~\cite{Binoth:1999sp,Denner:2005nn,Ossola:2006us}.
Moreover, no completely general libraries are available here (see e.g. LoopsTools~\cite{Hahn:2000jm} for public code).

Coming back to Fig.~\ref{pic:nlo-flowchart}, 
a characteristic feature of both contributions $d\sigma^{\rm real}$ and $d\sigma^{\rm virtual}$ 
is the presence of infrared (IR) divergencies due to soft and collinear regions in phase space.
The physical cross section $\sigma^{\rm NLO}_{2 \to n}$ being the sum of both is, of course, 
IR finite after absorbing the initial state collinear singularities 
into the PDFs by mass factorization (see e.g.~\cite{ellis:book} for details).
As we aim at parton level Monte Carlos to NLO accuracy with a 
flexible phase space integration allowing for kinematical cuts, 
the IR divergencies need to be treated accordingly.
Among the many proposed methods (see e.g.~\cite{Dixon:2007hh}), 
the so called dipole subtraction~\cite{Catani:1997vz,Catani:2002hc}
has emerged as a standard procedure.
Here, the cancellation of infrared singularities due to collinear partons or soft gluons 
is implemented ``locally'' by subtracting (over the entire phase space)
functions (the dipoles) 
that approximate the singularities of the real emission part.
Subsequently, the integrated dipoles are added to the virtual corrections. 
Employing dimensional regularization ($d = 4 - 2\epsilon$),
the master formula reads~\cite{Catani:1997vz,Catani:2002hc},
\begin{eqnarray}
  \label{eq:nlo-dipole}
  \sigma^{\rm NLO}_{2 \to n} &=&
  \int_{n+1} \left[ 
  \left( d\sigma^{\rm real} \right)_{\epsilon=0} - 
  \left( d\sigma^{\rm dipole} \right)_{\epsilon=0}
  \right] 
  +
  \int_{n} \left[ 
  d\sigma^{\rm virtual} + \int_1 d\sigma^{\rm dipole} 
  \right]_{\epsilon=0}
\, ,
\end{eqnarray}
which is understood to contain also the mass factorization of the remaining 
initial state collinear singularities. 
With Eq.~(\ref{eq:nlo-dipole}) one arrives at a finite partonic cross section $\sigma^{\rm NLO}_{2 \to n}$
which can be integrated numerically over the available phase space and convoluted with the PDFs. 
Let us stress however, that for any practical solution of Eq.~(\ref{eq:nlo-dipole}) 
speed and stability of the numerics are criteria of paramount importance. 
Quite often for instance, the programs for a particular scattering process need dedicated optimization.
Thus, presently there is a lot of room for technological progress with respect
to automatization and algorithms in order to fill the empty spaces in Tab.~\ref{tab:les-houches}.

\subsection{New theory developments}
One specific direction of research in theory during the past few years has
been towards new analytic techniques to calculate gauge theory amplitudes. 
The specific focus has been on a recursive approach in which all intermediate quantities 
are on-shell and hence gauge invariant (see~\cite{Bern:2007dw} for a recent review).
As a matter of fact, techniques for computing tree amplitudes recursively 
are well established since a number of years~\cite{Berends:1987me}.
Moreover, it has been realized that an efficient management of the quantum
numbers for a given scattering amplitude reduces the computational complexity by far.
The known methods include so-called color ordering, the use of helicity amplitudes and 
a decomposition of QCD amplitudes exploiting effective supersymmetry (SUSY) 
(see e.g.~\cite{Dixon:1996wi}).
In addition, factorization properties of amplitudes in the soft and collinear limits
serve as a strong check (see for instance~\cite{Catani:1997vz,Dixon:1996wi}).

In a helicity basis, amplitudes for scattering processes are classified
according to the number of `$\pm$'-states of the external partons.
The so-called maximal helicity violating (MHV) amplitude denotes the
configuration with the largest difference of `$+$' and `$-$'-states, e.g. 
$n-2$ for a $n$-gluon amplitude $A_n$ at tree level.
The tree level (color ordered) $n$-gluon MHV amplitude $A_n^{\rm tree}$  takes a particularly
simple and elegant form~\cite{Parke:1986gb}.
In terms of (Weyl) spinor inner-products 
$\spinorbra{j}{l} = \overline{u_-}(k_j) u_+(k_l)$ for massless 
Weyl spinors $u_\pm(k)$ of momentum $k$ we can write (see also e.g.~\cite{Mangano:1990by}),
\begin{eqnarray}
  \label{eq:mhv-parke-taylor}
  A_n^{\rm tree}(1^-,2^-,3^+,\dots,n^+) = {\rm i}
  {
    \spinorbra{1}{2}^4 \over
    \spinorbra{1}{2}\, \spinorbra{2}{3}\, \dots \, \spinorbra{n}{1}
  }  \, ,
\end{eqnarray}
which is in a certain sense ``all-order'' information, because
Eq.~(\ref{eq:mhv-parke-taylor}) holds for any number $n$ of external gluons.

The importance of helicity amplitudes became again apparent upon applying twistor space methods~\cite{Witten:2003nn} 
and by realizing that tree amplitudes $A_n^{\rm tree}$ in gauge theories possess unique
analytic properties which become manifest, if considered as functions of
complex momenta $k$, i.e. under a shift $k \to k(z)$ for a complex valued parameter $z$. 
These analyticity properties of $A_n^{\rm tree}(z)$ can be turned into recursion relations for 
the case of gluons~\cite{Britto:2004ap,Britto:2005fq} 
as well as quarks and scalars~\cite{Bern:2007dw,Dixon:2004za,Badger:2004ty},
which allow the construction of $n$-point helicity amplitudes from on-shell $(n-2)$-point amplitudes. 
From $A_n^{\rm tree}(z)$ the physical amplitude at $z=0$, i.e. $A_n^{\rm tree}(0)$ 
can be reconstructed simply by exploiting Cauchy's theorem of complex analysis.
The key feature of complex kinematics is the fact that the three-parton
primitive MHV amplitude, e.g. Eq.~(\ref{eq:mhv-parke-taylor}) for $n=3$, does
not vanish for on-shell complex momenta $k(z)$, while it does so for real $k$.
For details of the complex shift $k \to k(z)$ we refer to~\cite{Britto:2005fq}.

\begin{figure}[htb]
  \begin{center}
  \includegraphics[bbllx=120pt,bblly=650pt,bburx=475pt,bbury=720pt,angle=0,width=15.0cm]{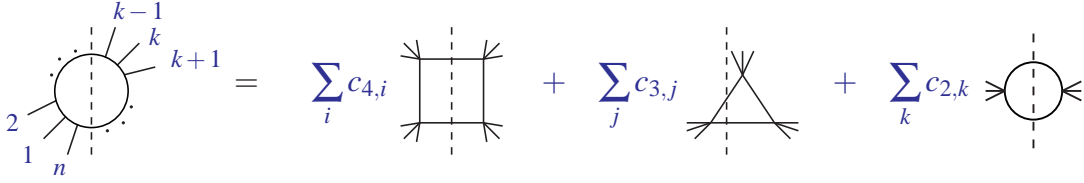}
\vspace*{-1mm}
\caption{ \small
\label{pic:unitarity}
 Schematic representation of the unitarity approach:
 Sewing of tree level amplitudes leads to the imaginary part of one-loop
 $n$-point amplitudes. The real part is determined subsequently from the
 analytic properties of the amplitudes 
 (adapted from Ref.~\cite{Mastrolia:2006vm}).
}
\vspace*{2mm}
  \end{center}
\end{figure}
The real quest, of course, has been in devising improved methods for the 
calculation of one-loop corrections to scattering amplitudes.
This has been a very active field of research over the past few years 
and it has been realized that unitarity methods provide the additional key ingredient here.
Unitarity is a fundamental concept in quantum field theory which manifests
itself for instance in the Cutkosky cutting rules for Feynman diagrams (related to the optical theorem).
In the computation of one-loop amplitudes, unitarity appears as fusing rules for amplitudes~\cite{Bern:2007dw,Bern:1994cg}. 
One aims at reconstructing the real part of a given one-loop amplitude from the imaginary one 
by sewing together tree level amplitudes.
To that end, one uses the fact that any one-loop amplitude can be expressed 
in a basis of scalar integral functions, i.e. boxes, triangles, and bubbles.
This is obvious from the standard reduction techniques, e.g.~\cite{Passarino:1979jh} 
and it is sketched schematically in Fig.~\ref{pic:unitarity}.
Unitarity cuts then allow to identify uniquely the contribution to the individual
integral (i.e. the coefficients $c_{4,i}$, $c_{3,j}$, $c_{2,k}$ in Fig.~\ref{pic:unitarity}) 
from its imaginary part, for example a triangle loop from calculating 
${\rm i} \pi \ln(-s) \to \ln^2(-s)$.

Generalized unitarity~(see e.g.~\cite{Britto:2004nc}) developed further the
idea of reconstructing the coefficients $c_{4,i}$, etc. 
by imposing quadruple cuts, which constrain 
all components of a given one-loop integral~(\ref{eq:tensor-int}).
As an upshot, all terms with logarithmic dependence are cut-constructible. 
One-loop QCD amplitudes, however, also contain rational (non-logarithmic) terms, 
which are rather difficult to derive and require substantially more effort.

To that end, as a further principle to organize the calculation a SUSY inspired decomposition 
of one-loop amplitudes has been very useful. 
For the amplitude $A_n^g$ with $n$ external gluons and a gluon circulating in the loop 
we can write down,
\begin{eqnarray}
  \label{eq:susy-decompostion}
  \displaystyle
  A_n^g &=&
  \underbrace{\Bigl(A_n^g + 4 A_n^f + 3 A_n^s\Bigr)}_{\displaystyle {\cal N}=4\, \mbox{SUSY}}
  \,\,\,\,
  -\,\,\,\, 4 \hspace*{-5mm}
  \underbrace{\Bigl(A_n^f + A_n^s\Bigr)}_{\displaystyle {\cal N}=1\, \mbox{chiral SUSY}}
  +
  \,\,\,\,
  \underbrace{{\phantom{\Bigl(}} A_n^s {\phantom{\Bigr)}}}_{\displaystyle {\cal N}=0\, \mbox{scalar}}
  \, ,
\end{eqnarray}
where the internal loop degrees of freedom are ordered in terms 
supersymmetric multiplets, i.e. the ${\cal N}=4$ multiplet (1 gluon, 4 Weyl fermions, 6 real scalars), 
the chiral ${\cal N}=1$ multiplet (1 Weyl fermion, 2 real scalars) 
and the ${\cal N}=0$ part with a complex scalar in the loop.
This has the great advantage that ${\cal N}=4$ and the ${\cal N}=1$
contributions are completely cut-constructible. For the rational part in the 
${\cal N}=0$ bit, more involved recursions have been developed.

Finally, the computation of the one-loop virtual corrections to the six gluon amplitude
has been completed through the effort of many groups~\cite{Bern:1994zx,%
Bern:1994fz,Bidder:2004tx,Bedford:2004nh,Bidder:2005ri,Britto:2005ha,Bern:2005cq,Bern:2005hh,Britto:2006sj,%
Berger:2006ci,Berger:2006vq,Xiao:2006vt} 
(see Tab.~\ref{tab:six-gluon}) and the analytic results have been confirmed 
for specific phase space points by a completely numerical evaluation~\cite{Ellis:2006ss}. 
Recently, the numerical approach has been developed further and shown to have promising potential~\cite{Giele:2008ve}.
We therefore expect that the NLO correction to the four-jet cross section at LHC are within sight.
\begin{table}
  \begin{center}
\begin{tabular}{|c||c|c|c|c|}
\hline
Amplitude  &${\cal N}=4$ & 
            ${\cal N}=1$ & 
            ${\cal N}=0$ &
            ${\cal N}=0$ \\
           &  & 
              & cut
              & rat \\
\hline
\hline
$- - + + + +$
           & %\href{http://arxiv.org/hep-ph/9403226}
             \cite{Bern:1994zx}
           & %\href{http://arxiv.org/hep-ph/9409265}
             \cite{Bern:1994fz}
           & %\href{http://arxiv.org/hep-ph/9409265}
             \cite{Bern:1994fz}
           & %\href{http://arxiv.org/hep-ph/0507005}
             \cite{Bern:2005cq}
\\
\hline
$- + - + + +$
           & %\href{http://arxiv.org/hep-ph/9403226}
             \cite{Bern:1994zx}
           & %\href{http://arxiv.org/hep-ph/9409265}
             \cite{Bern:1994fz}
           & %\href{http://arxiv.org/hep-th/0412108}
             \cite{Bedford:2004nh}
           & %\href{http://arxiv.org/hep-ph/0607014}
             %\href{http://arxiv.org/hep-ph/0607017}
             \cite{Berger:2006vq,Xiao:2006vt}
\\
\hline
$- + + - + +$
           & %\href{http://arxiv.org/hep-ph/9403226}
             \cite{Bern:1994zx}
           & %\href{http://arxiv.org/hep-ph/9409265}
             \cite{Bern:1994fz}
           & %\href{http://arxiv.org/hep-th/0412108}
             \cite{Bedford:2004nh}
           & %\href{http://arxiv.org/hep-ph/0607014}
             %\href{http://arxiv.org/hep-ph/0607017}
             \cite{Berger:2006vq,Xiao:2006vt}
\\
\hline
$- - - + + +$
           & %\href{http://arxiv.org/hep-ph/9409265}
             \cite{Bern:1994fz}
           & %\href{http://arxiv.org/hep-th/0410296}
             \cite{Bidder:2004tx}
           & %\href{http://arxiv.org/hep-ph/0507019}
             %\href{http://arxiv.org/hep-ph/0602178}
             \cite{Bern:2005hh,Britto:2006sj}
           & %\href{http://arxiv.org/hep-ph/0604195}
             \cite{Berger:2006ci}
\\
\hline
$- - + - + +$
           & %\href{http://arxiv.org/hep-ph/9409265}
             \cite{Bern:1994fz}
           & %\href{http://arxiv.org/hep-th/0502028}
             %\href{http://arxiv.org/hep-ph/0503132}
             \cite{Bidder:2005ri,Britto:2005ha}
           & %\href{http://arxiv.org/hep-ph/0602178}
             \cite{Britto:2006sj}
           & %\href{http://arxiv.org/hep-ph/0607017}
             \cite{Xiao:2006vt}
\\
\hline
$- + - + - +$
           & %\href{http://arxiv.org/hep-ph/9409265}
             \cite{Bern:1994fz}
           & %\href{http://arxiv.org/hep-th/0502028}
             %\href{http://arxiv.org/hep-ph/0503132}
             \cite{Bidder:2005ri,Britto:2005ha}
           & %\href{http://arxiv.org/hep-ph/0602178}
             \cite{Britto:2006sj}
           & %\href{http://arxiv.org/hep-ph/0607017}
             \cite{Xiao:2006vt}
\\
\hline
\end{tabular}
    \caption{ \small
      \label{tab:six-gluon}
      The analytic analytic computation of the one-loop QCD corrections to the 
      individual helicity configurations of the six-gluon amplitude in the
      decomposition of Eq.~(\ref{eq:susy-decompostion})
      as a community effort (adapted from Ref.~\cite{Mastrolia:2006vm}).
    }
  \end{center}
\end{table}

Let us end the discussion by mentioning a few directions for further development.
Clearly, the new techniques have to be employed in complete cross section 
calculations following the steps outlined in Sec.~\ref{sec:qcd@nlo}. 
Moreover, at LHC many processes of interest contain 
either gauge boson or bottom and top quarks (see Tab.~\ref{tab:les-houches}).
Thus, the formalism sketched above needs to be carried over to case of massive (colored) particles.
This requires a number of extensions, be it the helicity formalism or the 
methods for calculating massive one-loop integrals from generalized unitarity
cuts (see e.g.~\cite{Schwinn:2007ee,Britto:2006fc}).
We should also mention, that of course many other developments in theory have
pushed the precision frontier for QCD predictions further, be it for 
multi-loop calculations (NNLO and beyond), for resummations 
or simply for algorithms and tools. Unfortunately we could not touch those aspects.

\subsection{Complete NLO and NNLO results}
Let us conclude this Section by summarizing the state-of-the art for complete
NLO cross section predictions for many-particle production.
It has become clear in the preceeding discussion (see Tab.~\ref{tab:les-houches}) 
that current theory research is focused on processes with $2 \to 3$ and $2 \to 4$ potentially 
massive particles like top quarks, $W/Z$-bosons, etc..

A few outstanding results that have appeared in the last one or two years in
this respect address for instance Higgs production in the Standard Model. 
Here a number of reactions has been investigated at NLO, such as 
$pp \to H + 2$~jets via gluon fusion~\cite{Campbell:2006xx} (in the heavy top limit, see Sec.~\ref{sec:higgs}) 
and via weak interactions~\cite{Ciccolini:2007jr} 
as well as $pp \to H + 3$~jets in vector-boson fusion~\cite{Figy:2007kv}.
These results provide important information e.g. for the extraction of the Higgs coupling to vector bosons at LHC.
The production of vector bosons has been considered to NLO accuracy in QCD, 
e.g. for the reaction $pp \to VV + 2$~jets via vector boson fusion~\cite{Jager:2006zc,Jager:2006cp,Bozzi:2007ur},
for tri-boson production ($pp \to ZZZ,WWZ$)~\cite{Lazopoulos:2007ix,Hankele:2007sb} 
and for $pp \to WW + 1$~jet~\cite{Dittmaier:2007th,Campbell:2007ev}.
Especially, the latter case is an important background for Higgs production in the low
mass range ($H \to WW$) and subsequent semi-leptonic decay of the $W$-bosons. 
A largely complete list of NLO calculations for LHC processes including electroweak corrections 
and also a number of new physics signals with e.g. the SUSY QCD corrections 
has been given in Ref.~\cite{Buttar:2006zd}.

Some predictions for $pp \to 3$~particles are published as public codes. 
Many of these programs also provide continuous updates which are made available
to the community e.g. through the CEDAR project~\cite{cedar:hepcode}.
To mention a few explicitly, let us point out NLOJET++~\cite{Nagy:2001fj,Nagy:2003tz} 
which comes as a multipurpose C++ library for calculating jet cross sections at NLO, 
e.g. for three-jet rates in hadron collisions.
The program MCFM~\cite{Campbell:2000bg} calculates for instance the production of a gauge boson 
or a Higgs in association with jets at hadron colliders.
The PHOX family with DIPHOX~\cite{Binoth:1999qq} and JETPHOX~\cite{Aurenche:2006vj} 
deals specifically with hard QCD radiation of photons along with jets in hadron collisions, 
which is another important background for the low mass Higgs, e.g. in the di-photon mode ($H \to \gamma\gamma$). 
Predictions for the photon pair-production background including 
a resummation of the transverse momentum $p_t$ of the di-photon pair~\cite{Balazs:1997hv,Balazs:2007hr} 
have been subject of the most recent improvements in the program ResBos~\cite{Balazs:1997xd}.

\begin{figure}[htb]
  \begin{center}
    \includegraphics[width=5.0cm,angle=0]{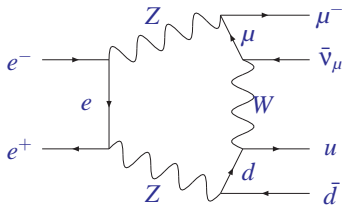}
\vspace*{-1mm}
\caption{ \small
\label{pic:hexagon}
Sample of Feynman diagram (hexagon with internal masses) for electroweak corrections to 
$e^+e^- \to 4$~fermions.
}
\vspace*{2mm}
  \end{center}
\end{figure}
NLO radiative corrections to scattering reactions with six external particles, i.e. $2 \to 4$ processes, 
constitute the current technological frontier. Benchmark results are the calculation of the
complete electroweak corrections to $e^+e^- \to 4$~fermions~\cite{Denner:2005es,Denner:2005fg}. 
This reaction involves at the loop level extremely difficult hexagon integrals with masses, see Fig.~\ref{pic:hexagon}. 
Of similar complexity are the NLO electroweak corrections to Higgs production
in association with a neutrino-pair, 
$e^+e^- \to \nu {\bar \nu} H H$, obtained by the GRACE group~\cite{Belanger:2002me,Boudjema:2005rk}. 
Also the NLO QCD corrections to the combined production of a top- and a bottom-pair, i.e. the process 
$\gamma\gamma \to t {\bar t} b {\bar b}$, are known~\cite{Lei:2007rv}.

Finally, there is of course demand for fully differential QCD predictions to NNLO for hadron collider processes.
As mentioned above, this scope has been achieved e.g. for the di-lepton pair production 
in Drell-Yan~\cite{Anastasiou:2005qj} or Higgs production in gluon fusion~\cite{Catani:2007vq} 
together with the parton evolution~\cite{Moch:2004pa,Vogt:2004mw}.
However, it remains a challenge for hadronic di-jet production, where large
statistics even with high-$p_t$ cuts is anticipated at LHC. 
The measurement of gluon jets would constrain for instance the gluon PDF at
medium and large $x$ and di-jet angular correlations are important observables 
in BSM searches for quark sub-structure. 
NNLO predictions for di-jets are likely to reduce the scale uncertainty and to improve 
the modeling of jets. 
Recent extensions of Tab.~\ref{tab:les-houches} also list the NNLO corrections 
to Higgs production in vector boson fusion, to top-pair production and 
to $V+1\mbox{jet}$, where $V \in \{\gamma,W^\pm,Z\}$.

Unfortunately, the calculation of NNLO cross sections is very difficult.
Although many (two-loop) virtual amplitudes are known since some years,
the cancellation of IR divergencies between virtual and real corrections
remains highly non-trivial and the numerical phase space integration very difficult, 
i.e. the NNLO equivalent of steps outlined in Fig.~\ref{pic:nlo-flowchart}.
This is a vast subject on its own, which we will not pursue further here 
(see e.g.~\cite{Dixon:2007hh} for a brief review).
Suffice it to say, that progress in this direction has been achieved only
recently for the differential distributions in the case of 
$e^+e^- \to 3$~jets~\cite{Gehrmann-DeRidder:2007bj,Gehrmann-DeRidder:2007hr}.

%
% -----------------------------------------------------------------------------
%
\section{Top quark production at LHC}
\label{sec:top}
Top quarks will be copiously produced at LHC. 
For the pair-production mode the collider will accumulate very high statistics of 
approximately $8 \cdot 10^6$ events with $t\bar{t}$-pairs 
with $10~{\rm fb}^{-1}$ per year 
in the initial low luminosity run~\cite{atlas:1999tdr2,cms:2006tdr}.
This data will allow for numerous measurements, e.g. of the top-mass, 
where the experiments aim at an accuracy of $\Delta m_t = {\cal O}(1) \mbox{GeV}$, 
and also for tests of the production and the subsequent decay mechanism 
including anomalous couplings and top-spin correlations (see e.g. Ref.~\cite{Quadt:2006jk}).
Top quark decay ($t \to Wb$) leads to very characteristic signatures 
allowing for event reconstruction in many channels through the observed
leptons, ($b$-flavored) jets and missing ${\slash E}_t$.
In general, top quarks make up large part of background for Higgs production or BSM searches.
Moreover, due to the large mass (currently $m_t = 170.9 \pm 1.1~(stat) + 1.5~(syst)$~GeV, see~\cite{tevewwg:2007bxa}) 
close to the scale of electroweak symmetry breaking top quarks play a
prominent role in many new physics scenarios (see e.g.~\cite{Frederix:2007gi}).

\subsection{Top quark pair-production}
\begin{figure}[htb]
  \begin{center}
    \includegraphics[width=4.65cm,angle=0]{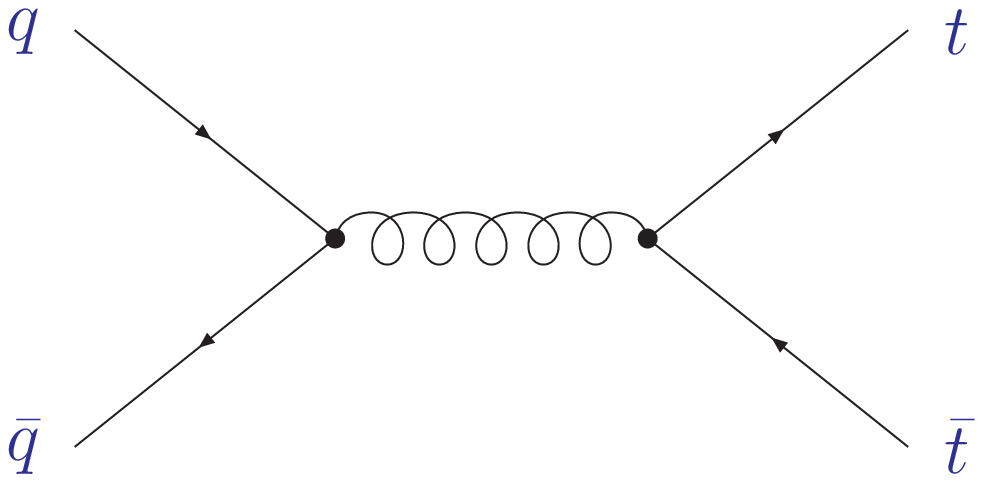}
    \includegraphics[width=11.5cm,angle=0]{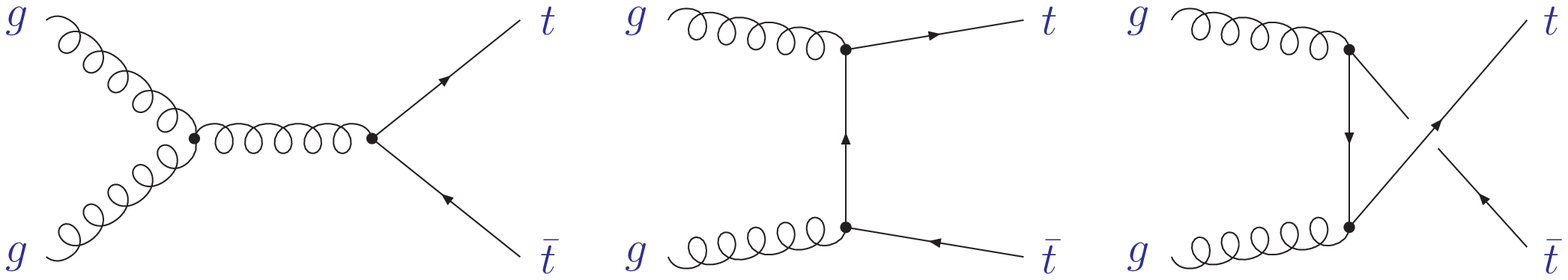}
\vspace*{-1mm}
\caption{ \small
\label{pic:ttbar-LO-diags}
Complete set of Feynman diagrams to leading order for the heavy quark pair-production in 
light quark annihilation, $q + {\bar q} \to t + {\bar t}$, (left) and in  
gluon fusion $g + g \to t + {\bar t}$ (three diagrams on the right).
}
\vspace*{2mm}
  \end{center}
\end{figure}

The hadronic heavy quark pair-production is known to NLO in QCD for many years~\cite{Nason:1988xz,Beenakker:1989bq} and 
it still serves as an example to illustrate many generic features of QCD corrections to the production of heavy colored particles,
see Fig.~\ref{pic:ttbar-LO-diags} for the corresponding Feynman diagrams to leading order.
Depending on the collider, i.e. Tevatron ($p{\bar p}$) or LHC ($pp$), 
the parton luminosities enhance the respective parton channels $q \bar q$
and $gg$. Thus, at Tevatron, $q \bar q$-annihilation saturates the
total cross section to ${\cal O}(90\%)$, while at LHC a similar dominance 
of gluon-fusion holds. 
The channels $qg ({\bar q}g)$ newly opening up at NLO contribute only ${\cal O}(1\%)$ at the scale $\mu = m_t$ 
at both colliders.
\begin{figure}[htb]
  \begin{center}
    \includegraphics[bbllx=50pt,bblly=280pt,bburx=355pt,bbury=500pt,angle=0,width=8.0cm]{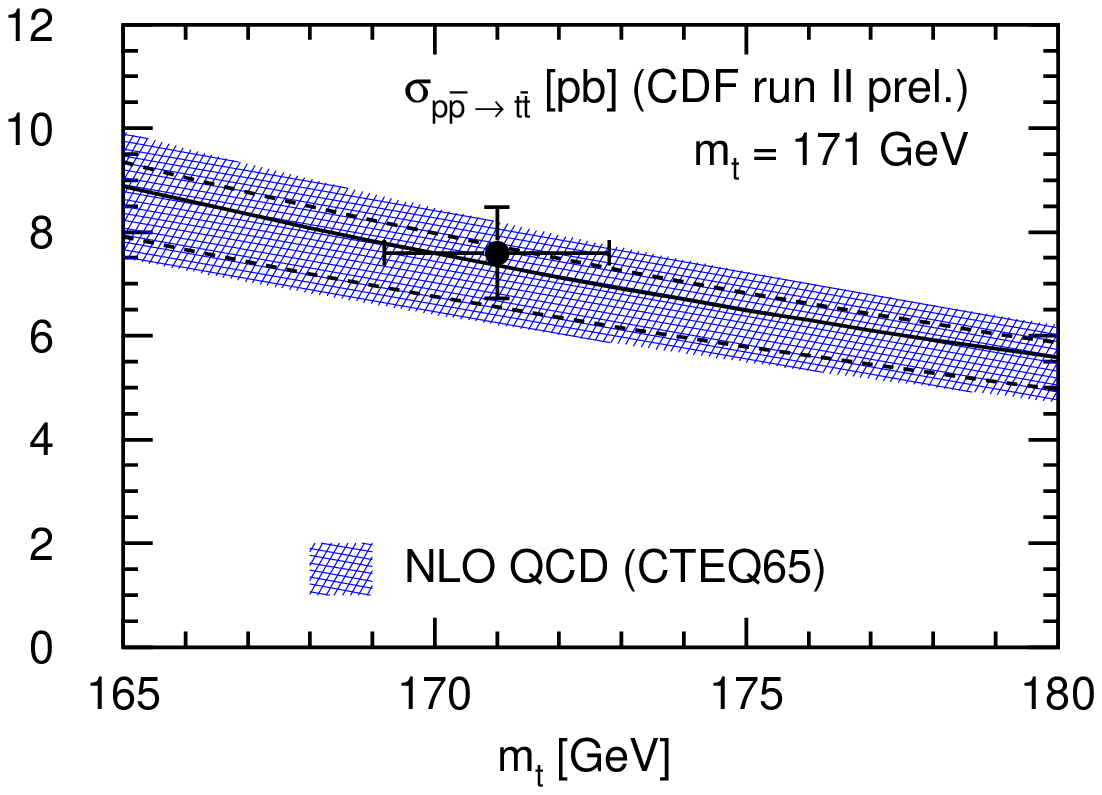}
\hspace*{0.25cm}
    \includegraphics[bbllx=50pt,bblly=280pt,bburx=355pt,bbury=500pt,angle=0,width=8.0cm]{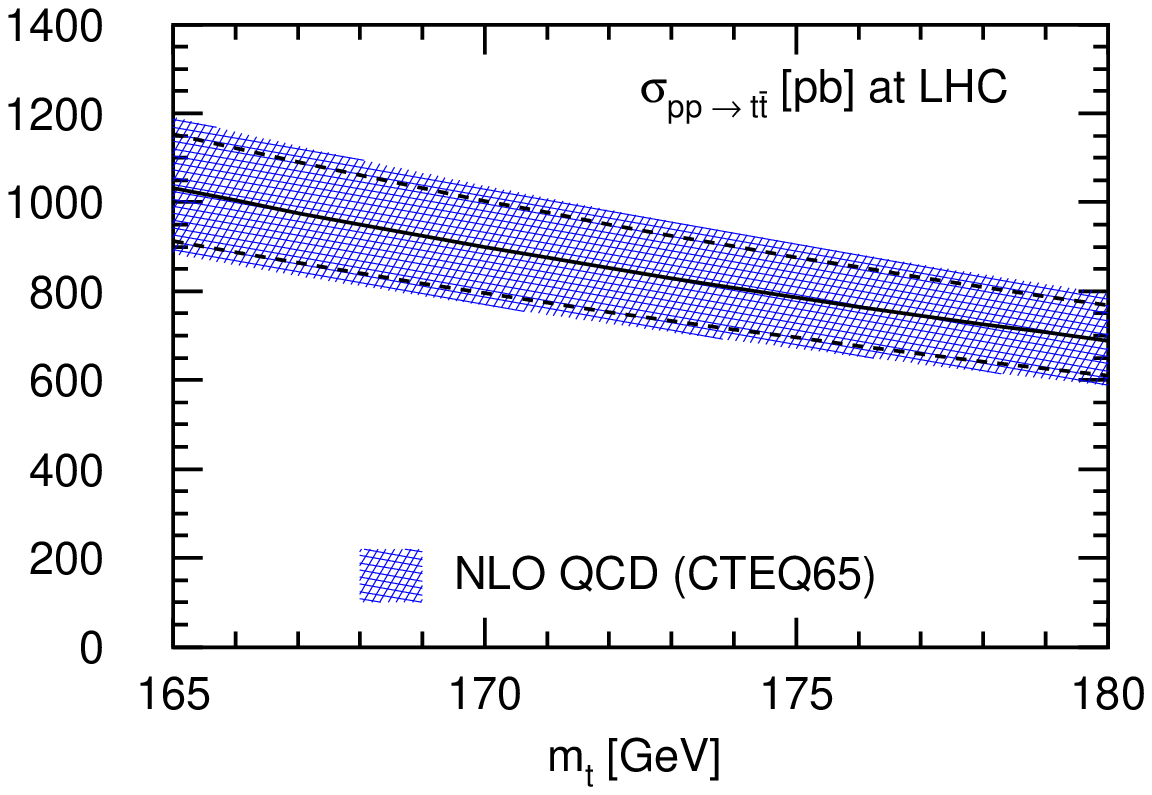}
\vspace*{-1mm}
\caption{ \small
\label{pic:ttbar-total}
The $t{\bar t}$ total cross section to NLO QCD as a function of $m_t$ for 
Tevatron at ${\sqrt{S}}=1.96$~TeV and CDF data~\cite{cdf:2006note} for $m_t=171$~GeV (left)
and LHC at ${\sqrt{S}}=14$~TeV (right).
The solid line is the central value for $\mu=m_t$, the dashed lower and upper
lines correspond to $\mu=2m_t$ and $\mu=m_t/2$, respectively.
The band denotes the additional PDF uncertainty of the CTEQ6.5 set~\cite{Tung:2006tb}.
}
\vspace*{2mm}
  \end{center}
\end{figure}
In Fig.~\ref{pic:ttbar-total} we show the NLO QCD predictions for the total cross-section of 
$t {\bar t}$-pair production at Tevatron and LHC.
The band denotes the scale variation in the usual range ($m_t/2 \ge \mu \ge 2 m_t$) as well as 
uncertainties related to the parton luminosity.
At Tevatron, the theory error budget at NLO in QCD is slightly asymmetric $+12 \%/-15 \%$ 
which breaks down to a scale uncertainty of $+5 \%/-10 \%$ and a PDF uncertainty of $+7 \%/-5 \%$. 
The latter one is due to the PDFs being sampled in the large-$x$ region, where
especially the gluon is poorly constrained.
At LHC, the total theory error is $15 \%$ which consists of a scale
uncertainty of $11 \%$ and a much smaller PDF uncertainty of $4 \%$.
Here, the cross section is sensitive to the gluon PDF in a range well
covered by HERA (see Fig.~\ref{pic:lhckin}).
Different sets of global PDFs agree within the given error bands, although it
should be pointed out that there can be sizable shifts in the central values.
For example, there is a $3 \%$ shift in the central value 
between the CTEQ6.5 and CTEQ6.6 sets~\cite{Nadolsky:2008zw} with 
correct heavy flavor treatment and the older set CTEQ6.1M 
(see also~\cite{Moch:2008tbp} for a recent discussion).

The theoretical prediction can be improved in specific kinematical regions.
Near threshold a Sudakov resummation can be performed~\cite{Kidonakis:1997gm,Bonciani:1998vc,Kidonakis:2001nj},
which stabilizes perturbative predictions if the $t{\bar t}$-pairs are produced
close to partonic threshold as for instance at Tevatron and, perhaps, to a lesser extent at LHC.
Further improvements of the theoretical accuracy need the NNLO QCD corrections, 
which are mandatory for a precision of better than ${\cal O}(10 \%)$ as envisaged by the LHC experiments.
First steps in this direction have been undertaken by evaluating the
interference of the one-loop QCD corrections~\cite{Korner:2008bn} 
and by deriving the virtual contributions to heavy-quark hadro-production at two loops 
in the ultra-relativistic limit $m^2 \ll s,t,u$~\cite{Czakon:2007ej,Czakon:2007wk}
based on a simple relation of massive and massless amplitudes in the limit $m \to 0$~\cite{Mitov:2006xs}
(see also the review~\cite{Moch:2007pj}).
A precise understanding of the kinematical region $m \to 0$ beyond NLO is of immediate
relevance also bottom-pair production over a large kinematical range and heavy
flavor production at large $p_t$. 

Top quark mass determinations at LHC are usually planned to proceed through direct
reconstruction because the theoretical accuracy of the total cross section is presently
insufficient, see Fig.~\ref{pic:ttbar-total}.
An interesting alternative (see e.g.~\cite{cms:2006tdr,Ball:2007zza}) 
involves a mass measurement through $J/\psi$ final states 
from $b$-decays in the top quark decay chain ($t \to Wb$).
The $J/\psi$-reconstruction is supposed to give an accurate measurement of the $b$-quark 
momentum thanks to the relatively high mass of the meson. 
However, currently, it is also limited by our knowledge of the heavy-quark fragmentation 
the  perturbative description of which has been extended to NNLO 
in QCD only recently~\cite{Melnikov:2004bm,Mitov:2004du,Mitov:2006ic,Moch:2007tx}.

\begin{figure}[htb]
  \begin{center}
    \includegraphics[angle=90, width=0.475\textwidth]{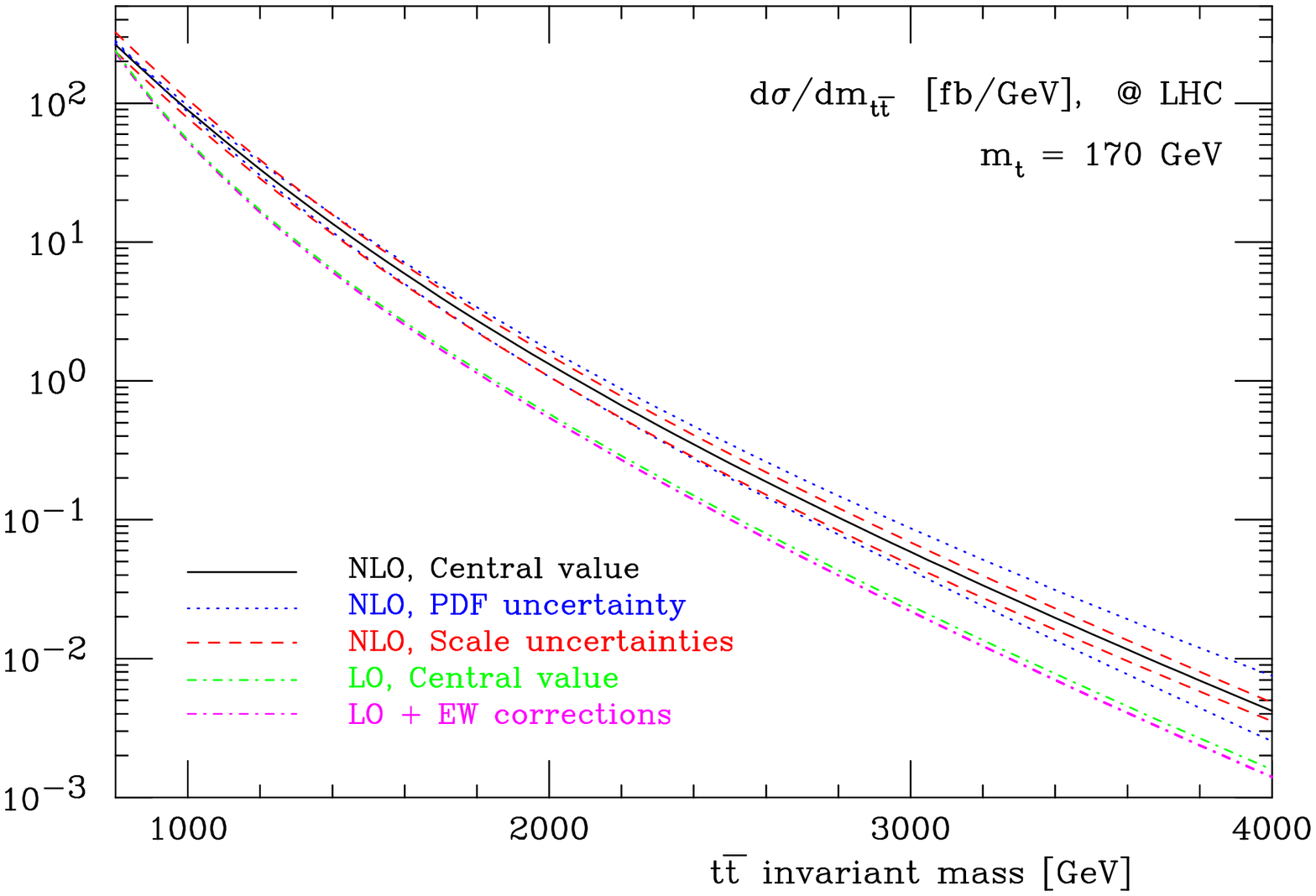}
    \includegraphics[angle=90, width=0.475\textwidth]{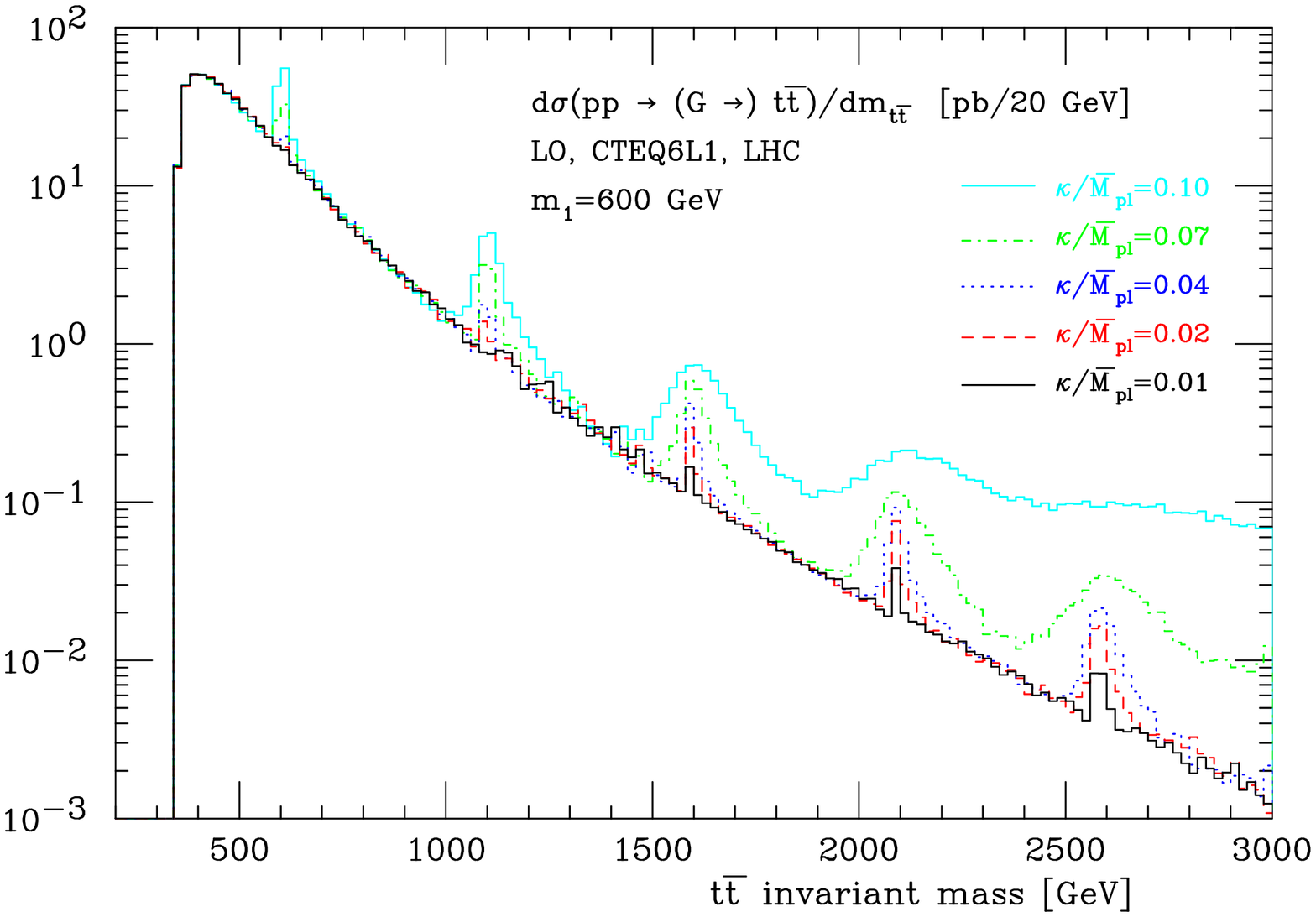}
\vspace*{-1mm}
\caption{ \small
\label{pic:ttbar-invmass}
Left: The $t\bar{t}$ invariant mass spectrum at LHC for $m_t=$ 170 GeV at NLO in QCD 
together with the scale (dashed) and the PDF (dotted) uncertainties 
for the CTEQ6.1M set.
Also plotted are predictions at LO in QCD (normalized to the NLO total cross section)
with (dark dash-dotted) and without (light dash-dotted) 
the NLO electroweak corrections for the CTEQ6L1 set 
(from Ref.~\cite{Frederix:2007gi}).
Right: The $t\bar{t}$ invariant mass spectrum at LHC including
$s$-channel graviton exchange and the effect of a couple
of Kaluza-Klein resonances in an extra dimensions model 
(from Ref.~\cite{Frederix:2007gi}, where the model parameters are specified).
}
\vspace*{2mm}
  \end{center}
\end{figure}
Other observables of experimental interest at LHC are differential distributions in the transverse momentum $p_t$ 
of an identified top quark, or in the invariant mass $m_{t{\bar t}}$ of the $t{\bar t}$-pair.
(see e.g.~\cite{Frederix:2007gi}).
Predictions for the latter are displayed in Fig.~\ref{pic:ttbar-invmass} (left) 
based on NLO in QCD as calculated e.g. with MCFM~\cite{Campbell:2000bg} 
and also including NLO electroweak corrections~\cite{Kuhn:2006vh}
for the CTEQ6.1 set~\cite{Pumplin:2002vw}. 
In the TeV-region for the invariant mass $m_{t {\bar t}}$ both the NLO QCD
corrections and also the electroweak radiative effects grow. 
The dominant theoretical errors however come from the scale and the PDF uncertainties. 
In particular the latter start to increase because the dominant contributions 
come again from the poorly known large-$x$ region.  

It has been pointed out~\cite{Frederix:2007gi} that the $m_{t{\bar t}}$-distribution 
also provides a window to new physics, where $s$-channel resonances may become visible. 
Fig.~\ref{pic:ttbar-invmass} (right) nicely illustrates the effect of graviton exchange 
in a model with one extra dimension compactified to a $\mathbf{S}^1/\mathbf{Z}_2$ orbifold
(see e.g.~\cite{Hooper:2007qk} and references therein).
Such a model leads to a tower of Kaluza-Klein modes 
giving rise to a series of resonances in the $t\bar{t}$ invariant mass spectrum, 
the details (peak position and width) of course, depending on the compactification
scale and the effective coupling $\kappa/{\bar M}_{pl}$.

\subsection{Single top quark production}
\label{sec:single-top}
The interest in single-top production at hadron colliders comes from a number of reasons.
It allows for studies of charged-current weak interactions of
the top quark and for a direct extraction of the CKM-matrix element $V_{tb}$.
Moreover, depending on the model under consideration, the cross section for 
single-top production acquires large corrections in BSM scenarios.
At Tevatron first evidence for single-top production has been found only
rather recently~\cite{Abazov:2006gd,Mitrevski:2008bx}.

\begin{figure}[htb]
  \begin{center}
    \includegraphics[width=4.4cm,angle=0]{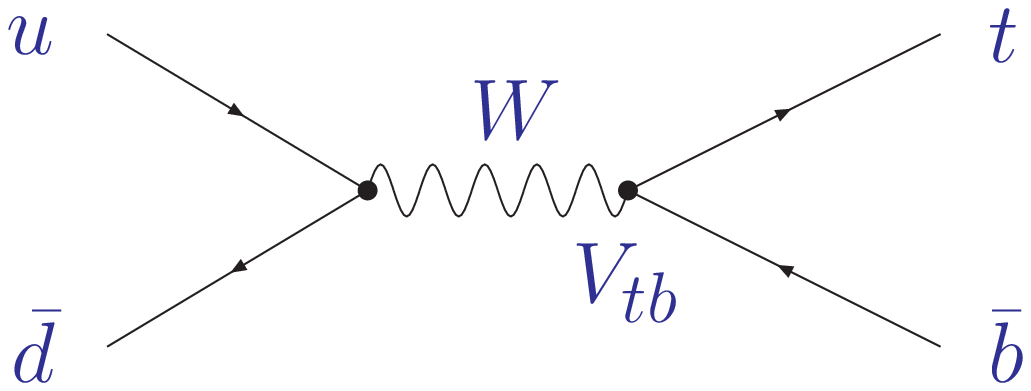}
    \includegraphics[width=8.2cm,angle=0]{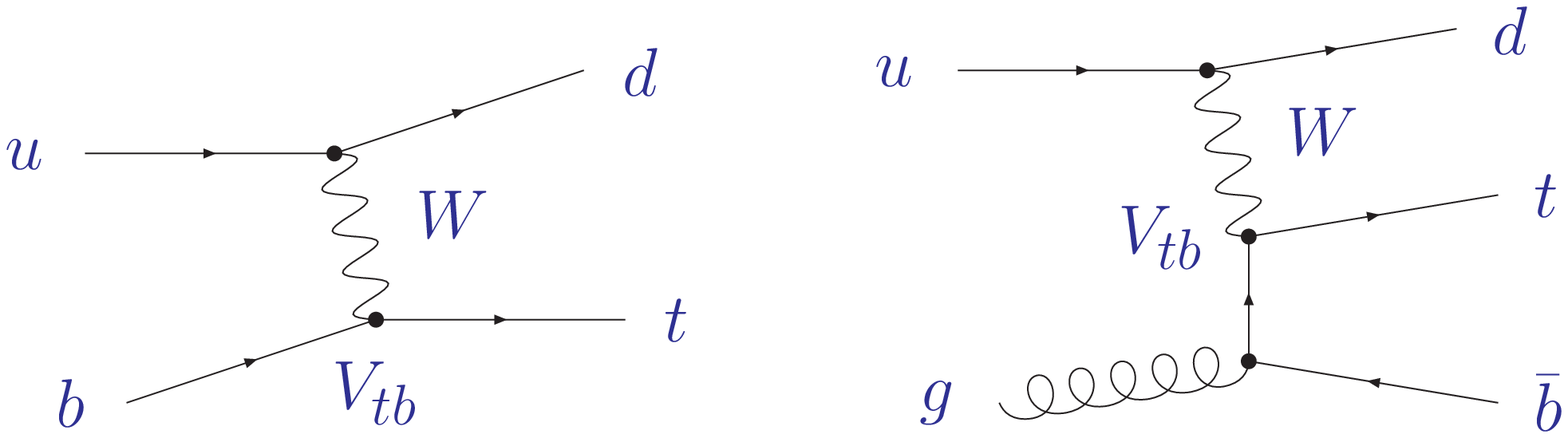}
    \includegraphics[width=3.65cm,angle=0]{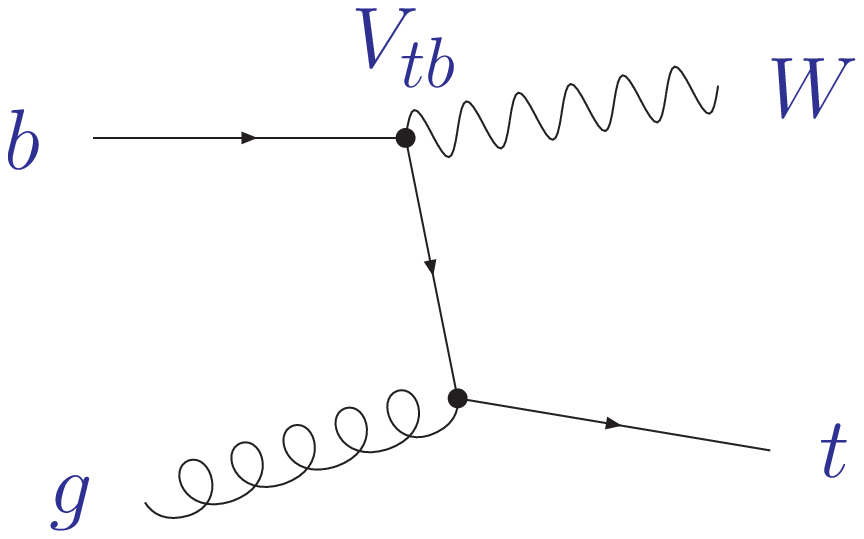}
\vspace*{-1mm}
\caption{ \small
\label{pic:singletop-LO-diags}
Sample of Feynman diagrams for single top-production.
Shown are the production in the $s$-channel (left), 
in the $b$-initiated $t$-channel (second from left), 
in the $g$-initiated $t$-channel (second from right)
and in the $bg$-channel (right). 
}
\vspace*{2mm}
  \end{center}
\end{figure}
The distinct channels for single-top production are displayed in Fig.~\ref{pic:singletop-LO-diags},
where samples of leading order Feynman diagrams are shown.
The $s$-channel mode (Fig.~\ref{pic:singletop-LO-diags}, left) proceeds
through the production of an off-shell $W$-boson and
subsequent decay ($W \to tb$). The initial state is proportional to the 
light flavor PDFs and the rate at LHC is relatively small.
The $t$-channel exchange of a virtual $W$-boson in the boson-gluon 
fusion mode (Fig.~\ref{pic:singletop-LO-diags}, second from right) on the other hand is the dominant production mechanism, both
at Tevatron and LHC. 
In the latter case, it is much enhanced due to the gluon PDF and subsequent 
splitting to a $b{\bar b}$-pair ($g \to b{\bar b}$), 
while the $b$-initiated $t$-channel process $ub \to dt$ itself
(Fig.~\ref{pic:singletop-LO-diags}, second from left) is suppressed by the numerically 
small bottom PDF. 
Of course, precise predictions at higher orders require a consistent matching of both processes, 
i.e. $g \to b{\bar b}$-splitting in the hard parton scattering and in the evolution~\cite{Sullivan:2004ie}.
For this main single-top production mode the NLO QCD corrections have also been
subject of the latest addition to MC@NLO~\cite{Frixione:2005vw}, 
so that NLO parton level calculations and showering are consistently combined.
Finally, there is $Wt$-production in the $bg$-channel (Fig.~\ref{pic:singletop-LO-diags}, right) 
being the second largest mode at LHC, 
but with negligible rates at Tevatron due to limited phase space.
The $bg$-channel is subleading in QCD (counting powers of the coupling constants) 
but again enhanced by the gluon luminosity.

Sensitivity to BSM models appears in different manifestations for the various channels of single-top production.
If one allows for anomalous couplings or flavor changing neutral currents, the $t$-channel contribution is altered significantly.
On the other hand, the existence of charged {\it "top-pions"},  excited Kaluza-Klein modes of the $W$-boson or, 
similarly, a $W^\prime$-boson would have impact on the $s$-channel.

\subsection{Top quark plus jet production}
The final example of this Section is concerned with top quark plus jet production
where, due to the high center-of-mass energy at LHC, also large statistics is expected. 
At the same time, the process $t{\bar t} +$~jets is an important background to Higgs or 
supersymmetry searches, so that experimental search strategies need to impose
kinematical cuts on the final state. 
At tree level, the cross sections have a really large scale dependence, 
which is why NLO QCD corrections are mandatory.
For instance, the process $t{\bar t} + 2$~jets entered Tab.~\ref{tab:les-houches} as a specific 
background to $t{\bar t}H$, where the Higgs decays into a $b{\bar b}$-pair.
The NLO QCD corrections to the former would help to control the background uncertainty
due to a heavy flavor mistag in a $t{\bar t} + 2$~jets event.

\begin{figure}[htb]
  \begin{center}
    \includegraphics[width=8.0cm,angle=0]{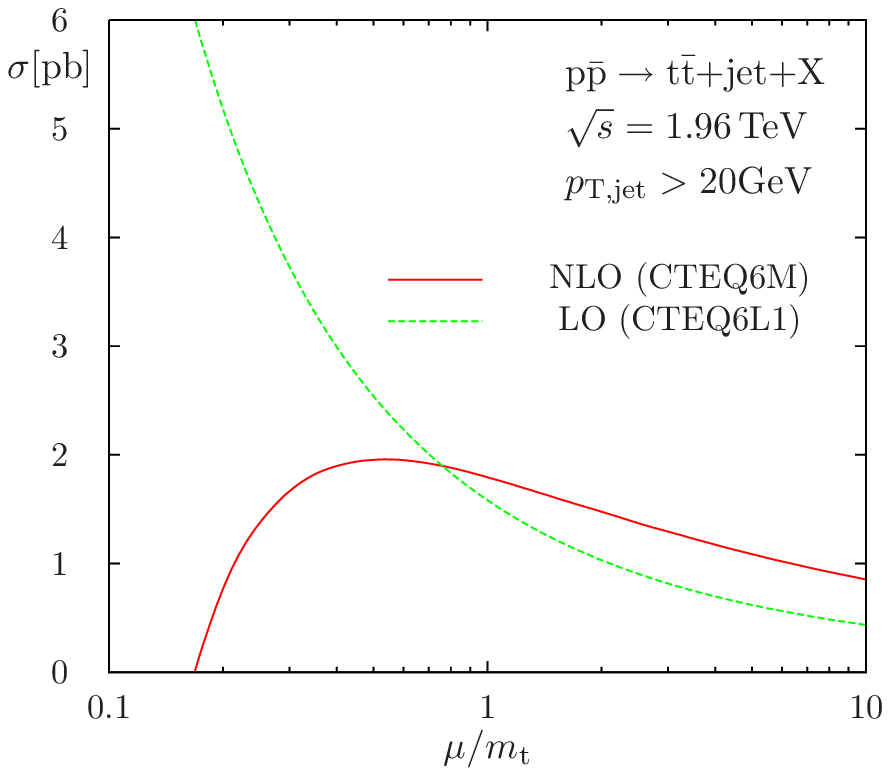}
    \includegraphics[width=8.0cm,angle=0]{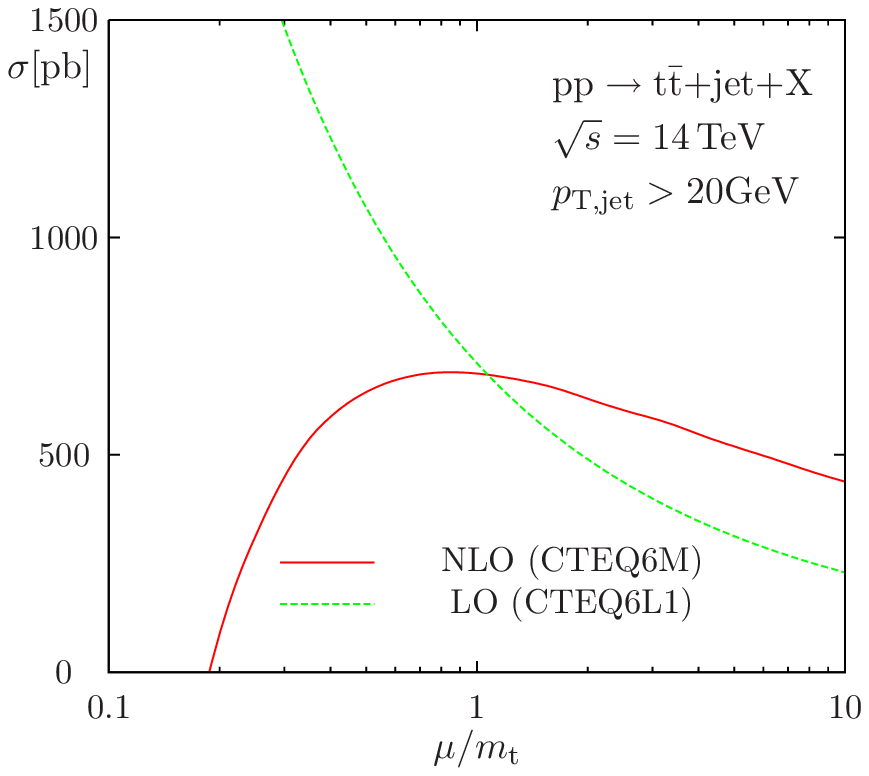}
\vspace*{-1mm}
\caption{ \small
\label{pic:ttbar-jet}
The scale dependence of the LO and NLO cross sections
for $t{\bar t} + 1$~jet production at the Tevatron (left) 
and at LHC (right) with renormalization and factorization scales identified, $\mu_r=\mu_f=\mu$ 
(from Ref.~\cite{Dittmaier:2007wz}).
}
\vspace*{2mm}
  \end{center}
\end{figure}
On the way to this challenge (six-leg processes being currently at the edge of technology), 
the process $t{\bar t} + 1$~jet was computed
recently to NLO QCD in an impressive state-of-the-art calculation~\cite{Dittmaier:2007wz}.
Fig.~\ref{pic:ttbar-jet} displays the much improved scale dependence and shows 
that the perturbative corrections are moderate for the nominal scale choice $\mu \simeq m_t$.
Clearly, it will be very interesting to see the NLO differential distributions 
for this reaction in the future and to compare them with LO predictions (e.g. from MadGraph)
combined with parton showers. 
In this way, one can assess how well the NLO predictions for jet-observables 
are modeled by the underlying partonic processes such as $gg \to t{\bar t}g$.

%
% -----------------------------------------------------------------------------
%
\section{Higgs production at LHC}
\label{sec:higgs}
\begin{figure}[htb]
  \begin{center}
    \includegraphics[bbllx=150pt,bblly=380pt,bburx=480pt,bbury=750pt,width=8.0cm,angle=0]{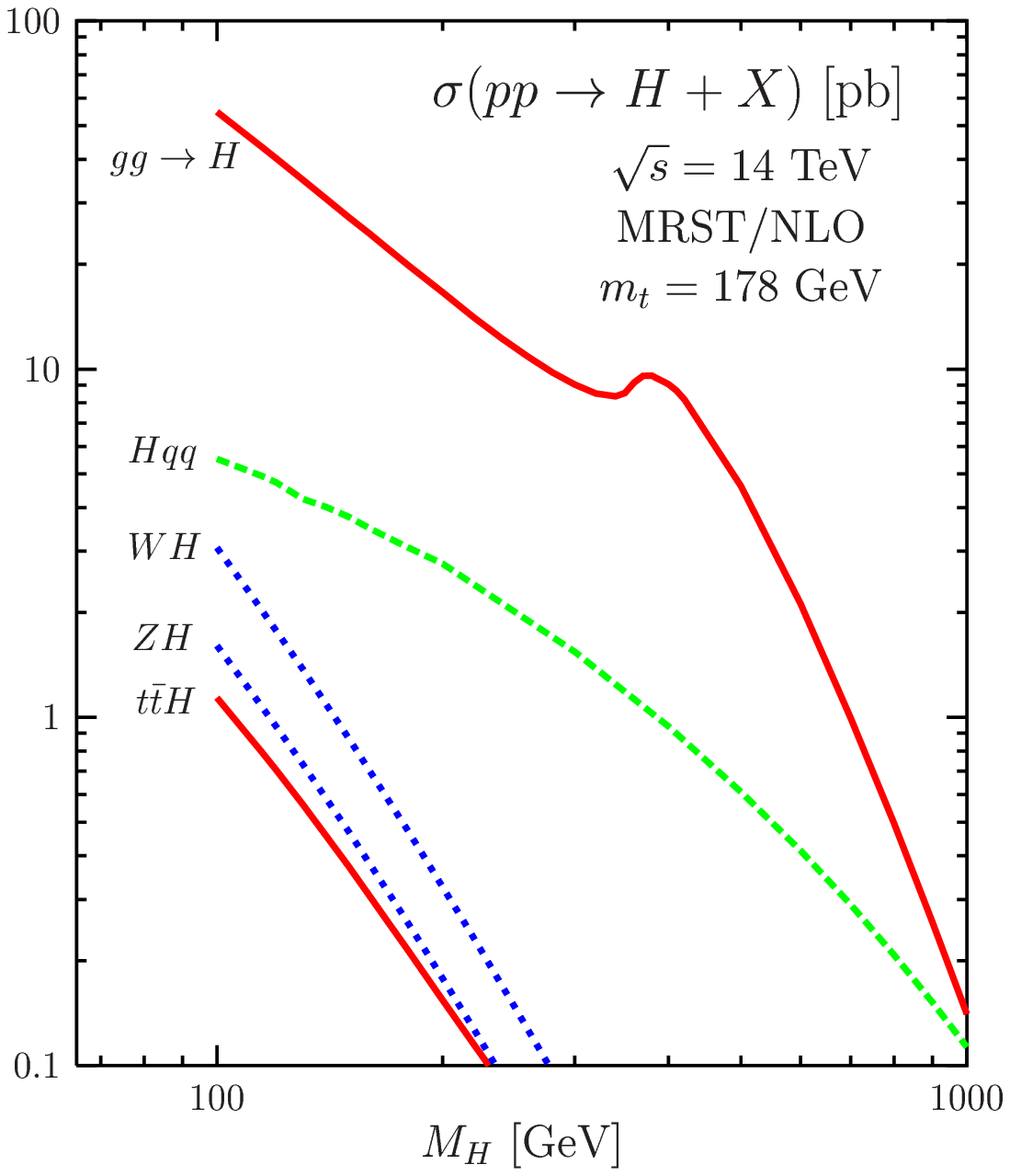}
    \includegraphics[bbllx=150pt,bblly=380pt,bburx=480pt,bbury=750pt,width=8.0cm,angle=0]{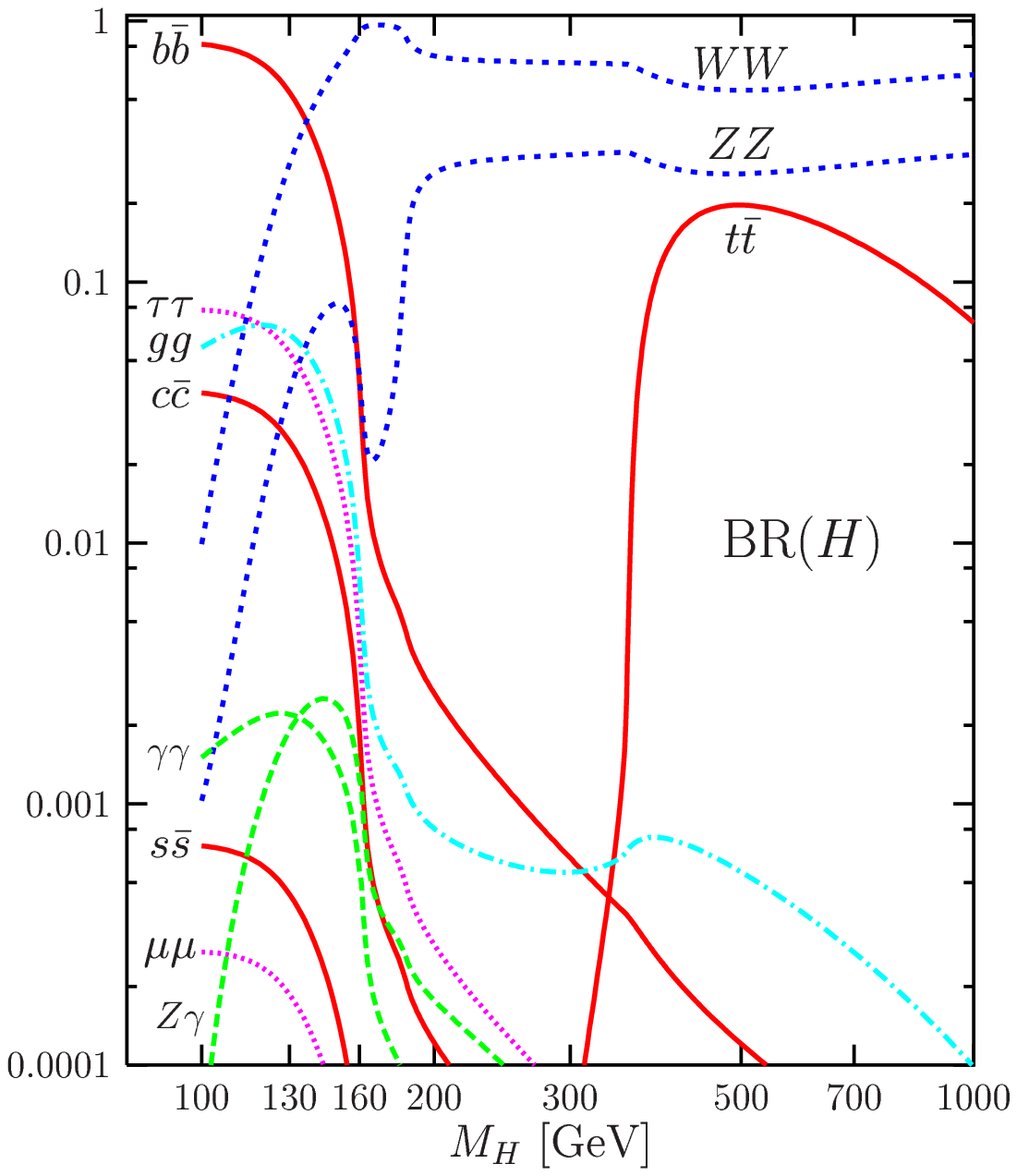}
\vspace*{-1mm}
\caption{ \small
\label{pic:Higgs-br-total}
Left: the total cross section for Higgs production at LHC at NLO in QCD.
Right: the branching ratios of Higgs boson decay 
(from Ref.~\cite{Djouadi:2005gi}).
}
\vspace*{2mm}
  \end{center}
\end{figure}

Let us conclude this review with a brief discussion of the flagship measurement to be conducted at LHC.
To start with, we illustrate in Fig.~\ref{pic:Higgs-br-total} for the Standard Model Higgs 
the dominant production modes (left) and the branching ratios for the decay (right) 
as a function of the Higgs mass.
The plotted values for the mass range up to $M_H = 1$~TeV, which is generally considered an upper
bound for the Standard Model Higgs due to triviality.
A lower bound on the Higgs mass has been established from direct searches at
LEP~\cite{Barate:2003sz,lep:eewg}, currently $M_H = 114.4$~GeV.
With the high luminosity and statistics of run-II the Tevatron experiments
currently conduct an active search for the Higgs as well (see~\cite{tev:eewg}).

Focusing on hard QCD aspects, we will limit ourselves to the production part 
(i.e. Fig.~\ref{pic:Higgs-br-total} on the left).
Depending on the Higgs mass, the signatures from the various decay modes
(gauge bosons, lepton pairs, quark pairs, etc.) define the 
experimental search strategy and, at the same time of course, the need to improve
predictions for the competing Standard Model processes as discussed in Sec.~\ref{sec:parton-crs}.
Unfortunately, this will not always be successful, as for example the fully hadronic modes will not be 
accessible for Higgs detection because of the huge QCD mult-jet background. 
However, due to the numerous channels we can hardly touch these aspects.
A very extensive discussion of Higgs physics at colliders can be
found e.g. in the recent Refs.~\cite{Djouadi:2005gi,Djouadi:2005gj}.
\begin{figure}[htb]
  \begin{center}
\includegraphics[width=4.0cm,angle=0]{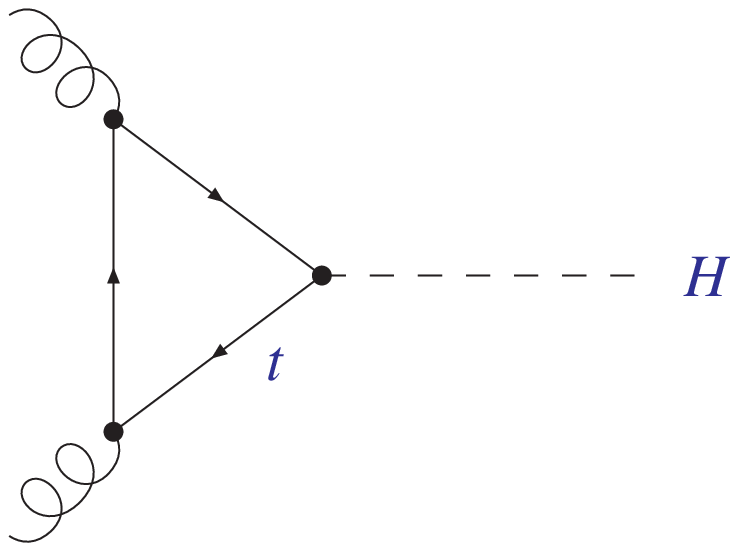}
\includegraphics[width=4.0cm,angle=0]{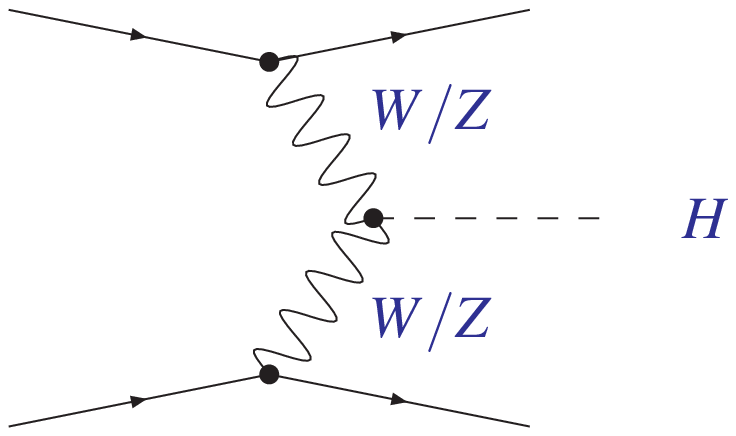}
\includegraphics[width=4.0cm,angle=0]{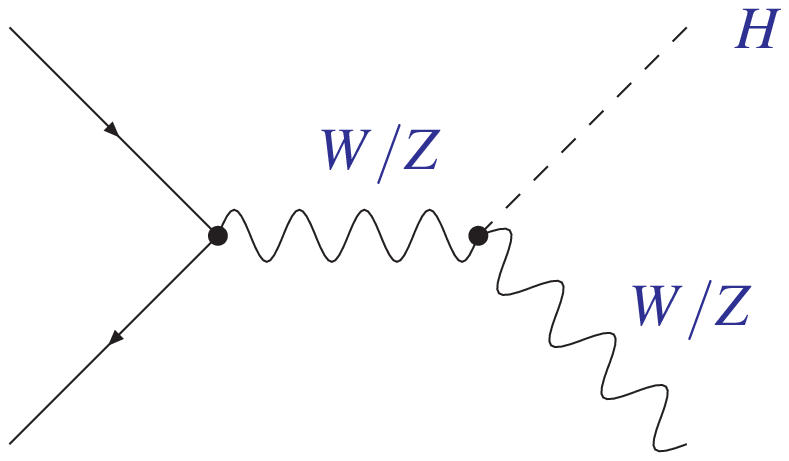}
\includegraphics[width=4.0cm,angle=0]{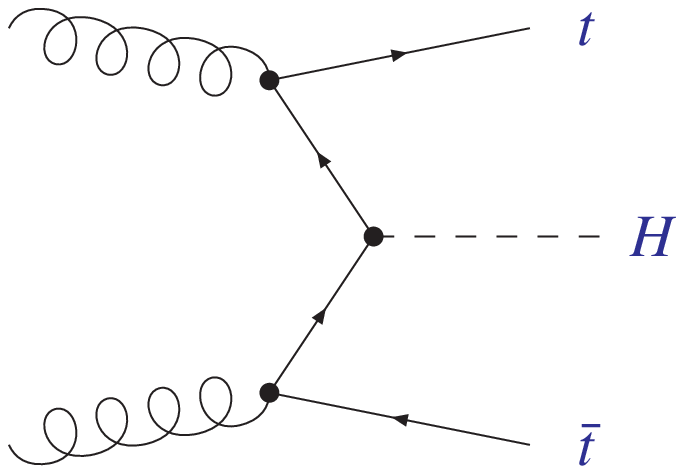}
\vspace*{-1mm}
\caption{ \small
\label{pic:higgs-LO-diags}
Sample of Feynman diagrams for the various modes of Higgs production.
Gluon fusion (left), weak vector-boson fusion (second from left),
Higgs-strahlung (second from right) and $t {\bar{t}}$H-channel (right).
}
\vspace*{2mm}
  \end{center}
\end{figure}
As can seen from Fig.~\ref{pic:Higgs-br-total} on the left, we have a clear hierarchy of channels and 
in Fig.~\ref{pic:higgs-LO-diags} we display samples of Feynman diagrams for these production modes.
Gluon fusion is induced via a heavy quark loop (Fig.~\ref{pic:higgs-LO-diags} left).
It has the largest rate for all values of the Higgs mass $M_H$ due to the large top-Yukawa coupling of the Higgs boson 
and the large gluon luminosity. 
In limit of a heavy top ($m_t \to \infty$) one can describe the interaction by 
an effective $ggH$-vertex upon integrating out the heavy quark in the loop,
which is a very good approximation also for finite $m_t$. 
QCD corrections for numerous observables in this channel have been
determined and we will high-light a few aspects below.

Weak vector-boson fusion proceeds via $q q \to q q H$ (Fig.~\ref{pic:higgs-LO-diags} second from left) 
and is mediated by $t$-channel gauge boson exchange.
It has the second largest rate, being dominated mostly by the $u, d$-quark PDFs 
and is proportional to the $WWH$ coupling. 
However, the signal identification for the three-body final state needs dedicated cuts 
on the final state jets (see e.g.~\cite{Rainwater:1997dg,Rainwater:1999sd}).
The characteristics of the latter are extremely important to 
discriminate the VBF signal from QCD backgrounds, for instance through 
forward jet-tagging and central jet-vetoing (see e.g.~\cite{Rainwater:1999gg}).
Higgs-strahlung in the channel $q {\bar q} \to W (Z) H$ (Fig.~\ref{pic:higgs-LO-diags} second from right) 
makes up for the third largest rate and has the same couplings as vector boson fusion.
The production mechanism requires a gauge boson from $q{\bar q}$-annihilation
so that the radiative corrections in QCD for Higgs-strahlung follow largely
from the corresponding ones in the Drell-Yan process and have been determined
to NNLO~\cite{Brein:2003wg}.

As another mode the associated Higgs production with heavy quarks 
has been discussed, for example $pp \to t {\bar{t}}H$ (Fig.~\ref{pic:higgs-LO-diags} right),
which is now known to NLO in QCD~\cite{Beenakker:2001rj,Beenakker:2002nc,Dawson:2002tg,Dawson:2003zu}.
At LHC the process is driven by the gluon luminosity but the rate drops quickly for larger Higgs masses and 
the phase space becomes too small already for values of $M_H \simeq 180$~GeV.
In addition, for lower Higgs masses the final state from $pp \to t {\bar{t}}H$
has a large Standard Model background (see Tab.~\ref{tab:les-houches}) 
which will be difficult to suppress.
The process $pp \to b {\bar{b}}H$ even has a slightly larger rate at LHC for 
$M_H \le 300$~GeV, but the final state with $b$-jets is overwhelmed by background.
Generally, Higgs couplings to bottom quarks are more important in the extended
Higgs sector of the MSSM (see e.g.~\cite{Djouadi:2005gj}).
Finally, the production of Higgs boson pairs ($pp \to HH + X$) at LHC still has rates accessible 
to measurements for low Higgs masses, however we will not discuss these
processes further. Cross sections for the production of three or more
Higgs bosons at LHC are too small~\cite{Djouadi:2005gi}.

\begin{figure}[htb]
  \begin{center}
    \includegraphics[width=12.5cm,angle=0]{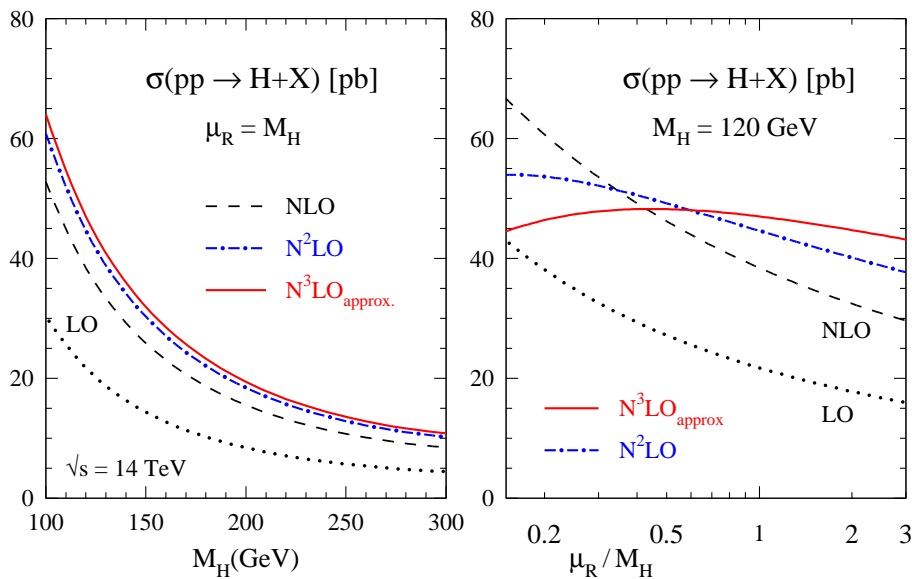}
\vspace*{-1mm}
\caption{ \small
\label{pic:higgs-nnlo}
Perturbative expansion of the total cross section for Higgs production
at the LHC. Shown are the dependence on the mass $M_H$ and the 
renormalization scale $\mu_{\rm r}$ (from Ref.~\cite{Moch:2005ky}).
}
\vspace*{2mm}
  \end{center}
\end{figure}

Let us, for the rest of this Section focus on the role of higher order QCD corrections for the 
Higgs signal in gluon fusion.  
The total cross section for gluon fusion in the heavy top limit is known
exactly to the NNLO in QCD~\cite{Harlander:2002wh,Anastasiou:2002yz,Ravindran:2003um}.
The radiative corrections yield a sizable $K$-factor of two or more as can be seen in Fig.~\ref{pic:higgs-nnlo} (left)
where we display the total cross section as a function of $M_H$ in the low mass range. 
Of course, the origin of this large effect is well understood. 
It is due to soft gluon emission, which makes up numerically for the bulk of the
perturbative corrections (see e.g.~\cite{Moch:2005ky} and references therein).
This fact has motivated the derivation of the complete soft N$^3$LO corrections in~\cite{Moch:2005ky} 
which are also plotted in Fig.~\ref{pic:higgs-nnlo} and illustrate nicely the property 
of apparent convergence of the perturbative expansion.
Another indicator in this respect is, as often stressed so far, the stability
under scale variation in a typical range, say $M_H \le \mu \le 2 M_H$.
This is shown in in Fig.~\ref{pic:higgs-nnlo} (right) where we display the 
total cross section as a function of the renormalization scale for a Higgs mass $M_H = 120$~GeV.
As the scale dependent terms are completely predicted by the lower order terms,
the curve denoted by N$^3$LO$_{\mbox{approx}}$ should be a very good
approximation of the exact three-loop result with a residual uncertainty
estimated of ${\cal O}(1-2\%)$ only.
\begin{figure}[htb]
  \begin{center}
    \includegraphics[width=8.0cm,angle=0]{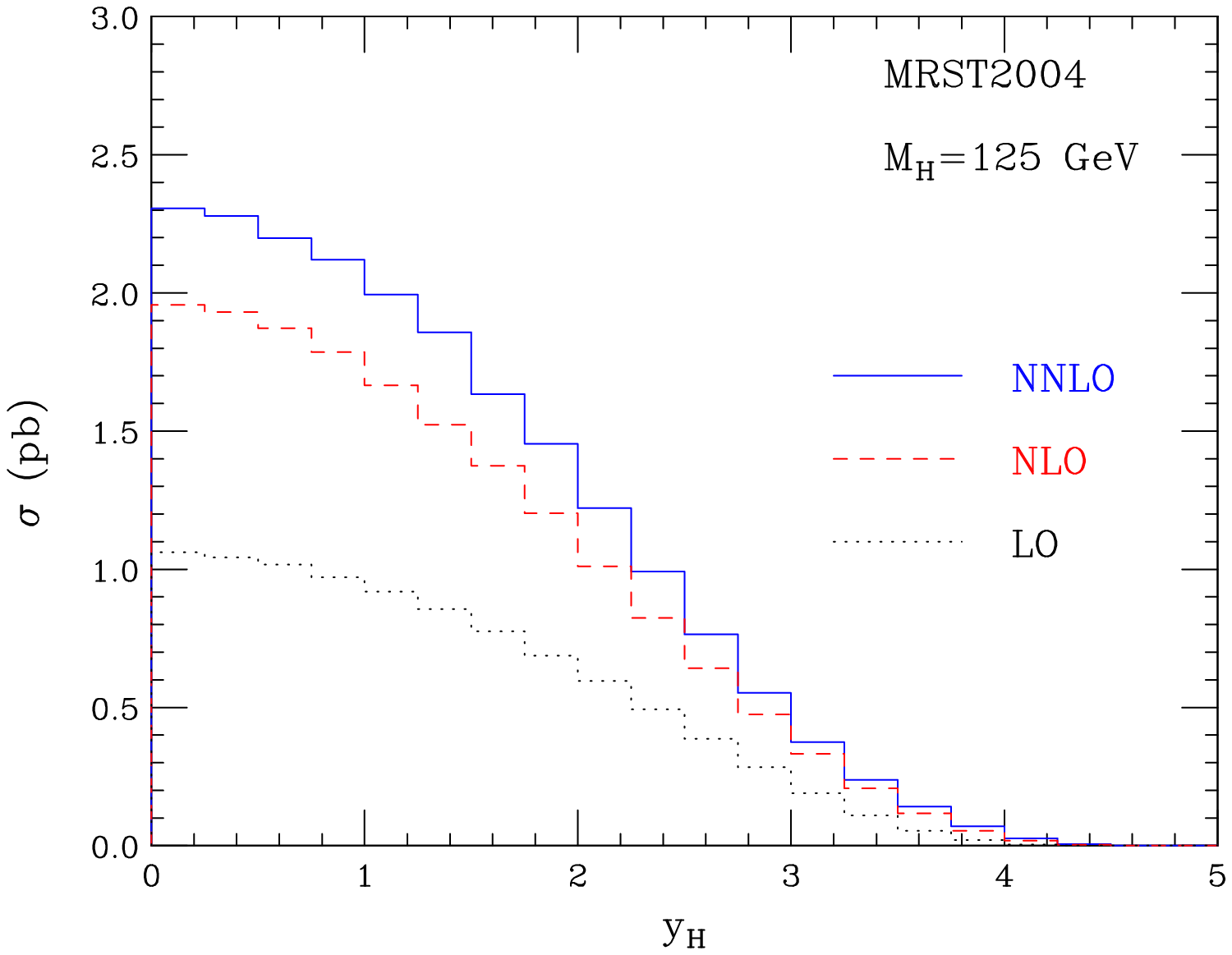}
    \includegraphics[width=8.0cm,angle=0]{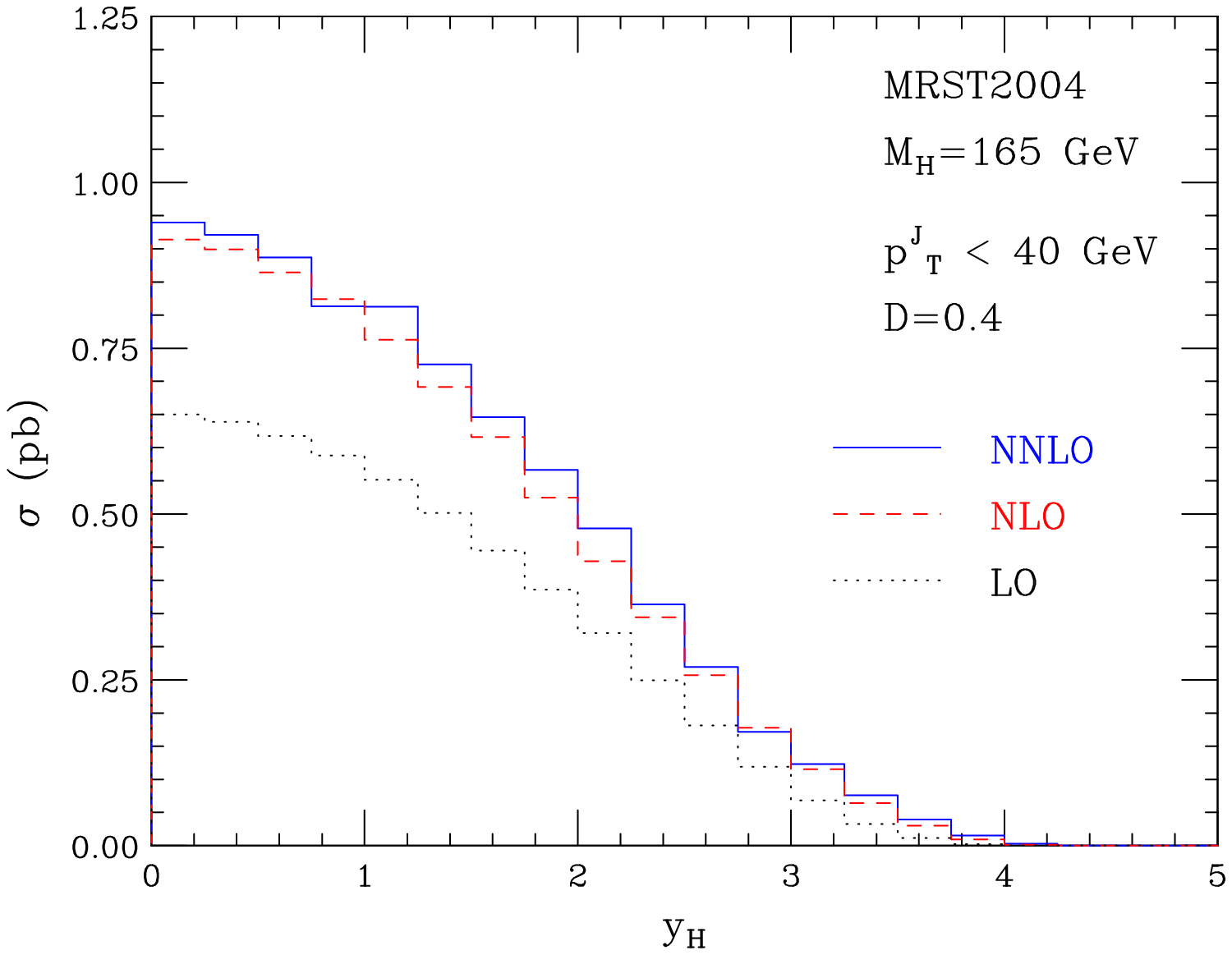}
\vspace*{-1mm}
\caption{ \small
\label{pic:higgs-nnlo-mc}
  Left: Higgs mass $M_h=125$~GeV, no cuts on $p_t$ of jets.
  Right: Higgs mass $M_h=165$~GeV and veto on jets with $p_t > 40$~GeV 
    ($k_t$ algorithm for jet reconstruction with jet size $D=0.4$)
    (from Ref.~\cite{Catani:2007vq}).
}
\vspace*{2mm}
  \end{center}
\end{figure}
Comparison with experimental data of course requires selection cuts on differential distributions. 
The latter are available for the gluon fusion channel including  QCD predictions up to NNLO 
and allow to study bin-integrated distributions (e.g. for the Higgs rapidity) with
subsequent Higgs decay in a variety of modes.
Presently the NNLO corrections in gluon fusion have been combined with the decay modes 
$H \to \gamma \gamma$ as well as  $H \to WW \to l\nu l\nu$ and 
$H \to ZZ \to 4 l$~\cite{Anastasiou:2005qj,Anastasiou:2007mz,Catani:2007vq,Grazzini:2008tf}. 
It is very interesting to study these higher order corrections under the impact of kinematical cuts on the
observed final state leptons, photons or the jet activity. 
Contrary to the findings for the total cross section, where higher order
corrections amount to $100\%$ or more, the effect of radiative
corrections in distributions is strongly reduced by the selection cuts.
We illustrate this for the Higgs rapidity
distribution as calculated with the parton level Monte Carlo program HNNLO~\cite{Catani:2007vq}.
In Fig.~\ref{pic:higgs-nnlo-mc} on the left no cuts on the $p_t$ of additional
jets are applied and the increase of the NNLO over the NLO corrections amounts to approximately
$20\%$ with some dependence on the rapidity. 
In contrast, in Fig.~\ref{pic:higgs-nnlo-mc} on the right 
all jets with $p_t \ge 40$~GeV have been vetoed, a situation typical 
when searching for the Higgs in the decay mode $H \to WW$ 
to suppress the $WW$ background from $t{\bar t}$-production.
As an upshot, the size of NNLO QCD radiative corrections is reduced to $5\%$.
Moreover, as expected, the NNLO QCD corrections improve significantly the stability 
under scale variation. 
First steps towards an assessment of the $WW$-background from QCD at NNLO 
have recently been made~\cite{Chachamis:2008yb}.
Also we remark that for the Higgs rapidity distribution even the soft N$^3$LO corrections have
been obtained and simple analytical formulae is available~\cite{Ravindran:2006bu}.

In summary we conclude that the rates for the Standard Model Higgs at LHC 
are reliably predicted by QCD. 
We have illustrated this for the gluon fusion channel where we have observed 
how higher order QCD predictions decrease the sensitivity to scale variations.
There exists a residual uncertainty of the cross section of a few per cent due
to the parton luminosity (see Sec.~\ref{pdfs-and-bsm}).
However, the gluon PDF is well constrained in the kinematical range and HERA data 
can be evolved to NNLO accuracy.
The study of differential distributions in particular with realistic
experimental cuts is an active field of ongoing research for all main
production modes (see Fig.~\ref{pic:higgs-LO-diags}) an we refer to the 
literature (see e.g.~\cite{Buttar:2006zd,Campbell:2006wx,Dixon:2007hh,Djouadi:2005gi}).

%
% -----------------------------------------------------------------------------
%
\section{Summary}
\label{sec:summary}

We have briefly reviewed the theoretical framework of QCD at hadron colliders.
Precision predictions for hard scattering cross sections rely on a detailed knowledge 
of the parton content of the proton and of the rates for the corresponding partonic subprocess.
We have given an overview of our current information on parton distributions 
including evolution to LHC energies.
For various Standard Model cross sections such as $W^\pm$/$Z$-boson, Higgs
boson or top quark production we have reported on the present status of perturbative QCD predictions.
Precision cross sections require the calculation of higher order corrections
for both, the signal and the background with massive particles and jets.
A lot of ongoing activity in this direction is concentrated on processes with multi-particle 
production and we have tried to give a snapshot of the technology involved,
e.g. the idea on-shell recursions for scattering amplitudes.
In summary, we have tried to convey the message that QCD theory is ready to meet 
the challenges of LHC.

In this review, we have mostly omitted details of specific hadronic final states, 
e.g. jet algorithms, $b$-quark ($b$-jet) production or aspects of $b$-quark fragmentation 
as well as parton showers in Monte Carlo simulations. 
We have also left out any discussion of resummation approaches meant to improve fixed order 
perturbation theory, be it threshold logarithms of Sudakov type or $\ln(p_t)$-terms in transverse momentum.
Finally, nothing has been said about the region of small-$x$ and forward physics 
at large rapidities (e.g. for diffractive production of Higgs bosons).
For all these remaining aspects as well as a broader coverage, 
the interested reader is referred 
to~\cite{atlas:1999tdr2,cms:2006tdr,Giele:2002hx,Dittmar:2005ed,Buttar:2006zd,Campbell:2006wx,Dixon:2007hh} 
and the numerous references therein.

QCD theory at hadron colliders is an extremely dynamical field at the moment, 
thus we expect that many specific issues will be improved or further clarified soon. 
The broad theoretical framework however, will certainly remain valid and 
hopefully, the present compilation has high-lighted the role of QCD 
in the era of LHC.

%
% ---------------------------------------------------------------------
%
\subsection*{Acknowledgments}
Feynman diagrams have been drawn with the packages 
{\sc Axodraw}~\cite{Vermaseren:1994je} and {\sc Jaxo\-draw}~\cite{Binosi:2003yf}.
The work has been supported in part by the Helmholtz Gemeinschaft under contract VH-NG-105.

\end{document}